\documentclass[12pt,times]{agujournal}
\usepackage{float}
\usepackage[bookmarks = true, bookmarksnumbered = true, pdfpagemode =None, pdfstartview = FitH, pdfpagelayout = SinglePage, colorlinks = true, urlcolor = blue, citecolor = blue]{hyperref}

\journalname{Geophysical Research Letters}

\begin{document}

%
%


\title{Jupiter's North Equatorial Belt expansion and thermal wave activity ahead of Juno's arrival}

%
%




\authors{L.N. Fletcher\affil{1}, G.S. Orton\affil{2}, J.A. Sinclair\affil{2}, P. Donnelly\affil{1}, H. Melin\affil{1}, J.H. Rogers\affil{3}, T.K. Greathouse\affil{4}, Y. Kasaba\affil{5}, T. Fujiyoshi\affil{6}, T.M. Sato\affil{7}, J. Fernandes\affil{8}, P.G.J. Irwin\affil{9}, R.S. Giles\affil{9}, A.A. Simon\affil{10}, M.H. Wong\affil{11}, M. Vedovato\affil{12}}

\affiliation{1}{Department of Physics \& Astronomy, University of Leicester, University Road, Leicester, LE1 7RH, UK.}
\affiliation{2}{Jet Propulsion Laboratory, California Institute of Technology, 4800 Oak Grove Drive, Pasadena, CA, 91109, USA.}
\affiliation{3}{British Astronomical Association, Burlington House, Piccadilly, London  W1J 0DU, UK.}
\affiliation{4}{Southwest Research Institute, San Antonio, Texas, USA.}
\affiliation{5}{Department of Geophysics, Tohoku University, Sendai, Japan.}
\affiliation{6}{Subaru Telescope, National Astronomical Observatory of Japan, National Institutes of Natural Sciences, 650 North A'ohoku Place, Hilo, Hawaii 96720, USA.}
\affiliation{7}{Institute of Space and Astronautical Science, Japan Aerospace Exploration Agency, 3-1-1, Yoshinodai, Chuo-ku, Sagamihara, Kanagawa 252-5210, Japan}
\affiliation{8}{California State University, Long Beach, CA, USA.}
\affiliation{9}{Atmospheric, Oceanic \& Planetary Physics, Department of Physics, University of Oxford, Clarendon Laboratory, Parks Road, Oxford, OX1 3PU, UK}
\affiliation{10}{NASA Goddard Space Flight Center Solar System Exploration Division (690) Greenbelt, MD 20771, USA}
\affiliation{11}{University of California at Berkeley, Astronomy Department Berkeley, CA 947200-3411, USA}
\affiliation{12}{JUPOS Team, Unione Astrofili Italiani}




\correspondingauthor{Leigh N. Fletcher}{leigh.fletcher@leicester.ac.uk}





\begin{keypoints}
\item Jupiter's North Equatorial Belt expanded northward in 2015/16 as part of its 3-5 year activity cycle.
 
\item The expanded sector warmed, removing light-colored aerosols to reveal darker colors at depth.
 
\item Significant thermal waves were identified in the troposphere and stratosphere at the time of the expansion.
 
\end{keypoints}

%
%


\begin{abstract}

The dark colors of Jupiter's North Equatorial Belt (NEB, $7-17^\circ$N) appeared to expand northward into the neighboring zone in 2015, consistent with a 3-5 year cycle. Inversions of thermal-IR imaging from the Very Large Telescope revealed a moderate warming and reduction of aerosol opacity at the cloud tops at $17-20^\circ$N, suggesting subsidence and drying in the expanded sector.  Two new thermal waves were identified during this period: (i) an upper tropospheric thermal wave (wavenumber 16-17, amplitude 2.5 K at 170 mbar) in the mid-NEB that was anti-correlated with haze reflectivity; and (ii) a stratospheric wave (wavenumber 13-14, amplitude 7.3 K at 5 mbar) at $20-30^\circ$N.  Both were quasi-stationary,  confined to regions of eastward zonal flow, and are morphologically similar to waves observed during previous expansion events.

\end{abstract}

%

\section{Introduction}
\label{intro}

Jupiter's banded appearance can exhibit dramatic global-scale variability, with the dark belts expanding, contracting, and sometimes whitening for many months at a time \citep[e.g.][and references therein]{95rogers, 96sanchez, 17fletcher_seb}.  These are the visible manifestations of thermal, chemical and aerosol changes throughout Jupiter's upper troposphere, which can be best diagnosed at thermal-infrared wavelengths.  As NASA's Juno spacecraft lacks instrumentation at these wavelengths, a campaign of Earth-based observations is underway to characterize Jupiter's thermal structure.  Here we report on the identification of a poleward expansion of Jupiter's North Equatorial Belt (NEB) and significant thermal wave activity in observations acquired in the months before Juno's arrival.  

The NEB is a broad, brown, cyclonic belt typically residing between eastward and westward jets at $6.8^\circ$N and $17.1^\circ$N, respectively (all latitudes are planetographic).  Visible observations have shown that the northern edge exhibits regular northward expansion events \citep{95rogers}, causing a decrease in reflectivity (darkening) of the southern edge ($17-21^\circ$N) of the white North Tropical Zone (NTrZ) with a 3-to-5-year period since 1987 \citep{17rogers}.  Supplementary Text S1 and Figure \ref{figS1} show details of seven NEB expansion events recorded since 1987.  These events either start from a single `bulge' of dark material encroaching into the NTrZ, or occur simultaneously at multiple longitudes, infilling and spreading over all longitudes over $\sim$5-7 months (Text S1).  The newly-darkened NTrZ then whitens over a 1-3 years, such that the edge of the brown belt appears to recede southwards to the normal $7-17^\circ$N range.  Infrared observations of the 2009-11 event demonstrated that the darkening of the NTrZ corresponded to a removal of aerosol opacity at 5 $\mu$m \citep{17fletcher_seb}.  The cyclic nature of the NEB activity suggested that a new expansion phase would occur in 2015-16, which we characterize here.


Jupiter's NEB and NTrZ also host wave phenomena.  A long-lived pattern of cloud-free `hotspots' and associated plumes has been observed on the NEB southern edge \citep[][and references therein]{13choi}, manifesting an equatorially-trapped Rossby wave \citep{90allison, 00showman, 05friedson}.  The interior of the NEB is characterized by upper-tropospheric thermal wave activity \citep[e.g.,][]{90magalhaes, 94orton, 04rogers, 06li, 16fisher}.  Stratospheric waves in the $20-30^\circ$N region were also reported \citep{91orton, 16fletcher_texes}.  A chronology of these previous wave detections is presented in Supplemental Text S2.  

\section{Earth-based thermal observations}

Narrow-band 8-25 $\mu$m thermal imaging of Jupiter (Table S1) was acquired in January 2016 by the COMICS instrument \citep{00kataza} on the Subaru Telescope, and between February and August 2016 by the VISIR instrument \citep{04lagage} on ESO's Very Large Telescope (VLT, Fig. \ref{figS2}).  These 8-m primary mirrors provide diffraction-limited spatial resolutions of 0.25-0.80" (700-2300 km at jovian opposition).  Full details of both instruments and the strategy for Jupiter observations are given by \citet{17fletcher_seb}.  \citet{09fletcher_imaging} describe image-reduction procedures: images were despiked and cleaned for bad pixels, geometrically registered by fitting the planetary limb, cylindrically reprojected and calibrated via comparison to Cassini Composite Infrared Spectrometer observations.  This time series was supplemented in January and May 2016 by spectroscopic mapping at 17.1 and 8.0 $\mu$m (at spectral resolutions of 5800 and 12400, respectively) using the TEXES instrument on NASA's Infrared Telescope Facility (IRTF) \citep{02lacy}, albeit at lower diffraction-limited spatial resolution from IRTF's smaller 3-m primary mirror.  \citet{16fletcher_texes} give full details of the spectral scanning technique.  Images and scans acquired over two or more consecutive nights were combined to create near-global maps of Jupiter.


Radiance maps in eight VISIR filters were stacked to form spectral cubes for inversion using an optimal-estimation retrieval algorithm \citep{08irwin}.  This allows estimates of the spatial variability of stratospheric temperatures near 5 mbar from CH$_4$ emission at 7.9 $\mu$m, 100-600 mbar tropospheric temperatures from 13-24 $\mu$m observations of the H$_2$ and He collision-induced continuum, and tropospheric aerosols near 600-800 mbar from 8-12 $\mu$m observations of NH$_3$ absorption bands.  With only eight photometric points defining each spectrum, the inversion suffers from limited vertical resolution and some degeneracies between retrieved parameters (particularly for high pressures).  Details of the inversion methodology and sources of spectral line data are given by \citet{09fletcher_imaging}.

\section{Results and Discussion}

\subsection{Chronology of the 2015-16 NEB Expansion}

\begin{figure}[h]
\centering
\includegraphics[width=18cm]{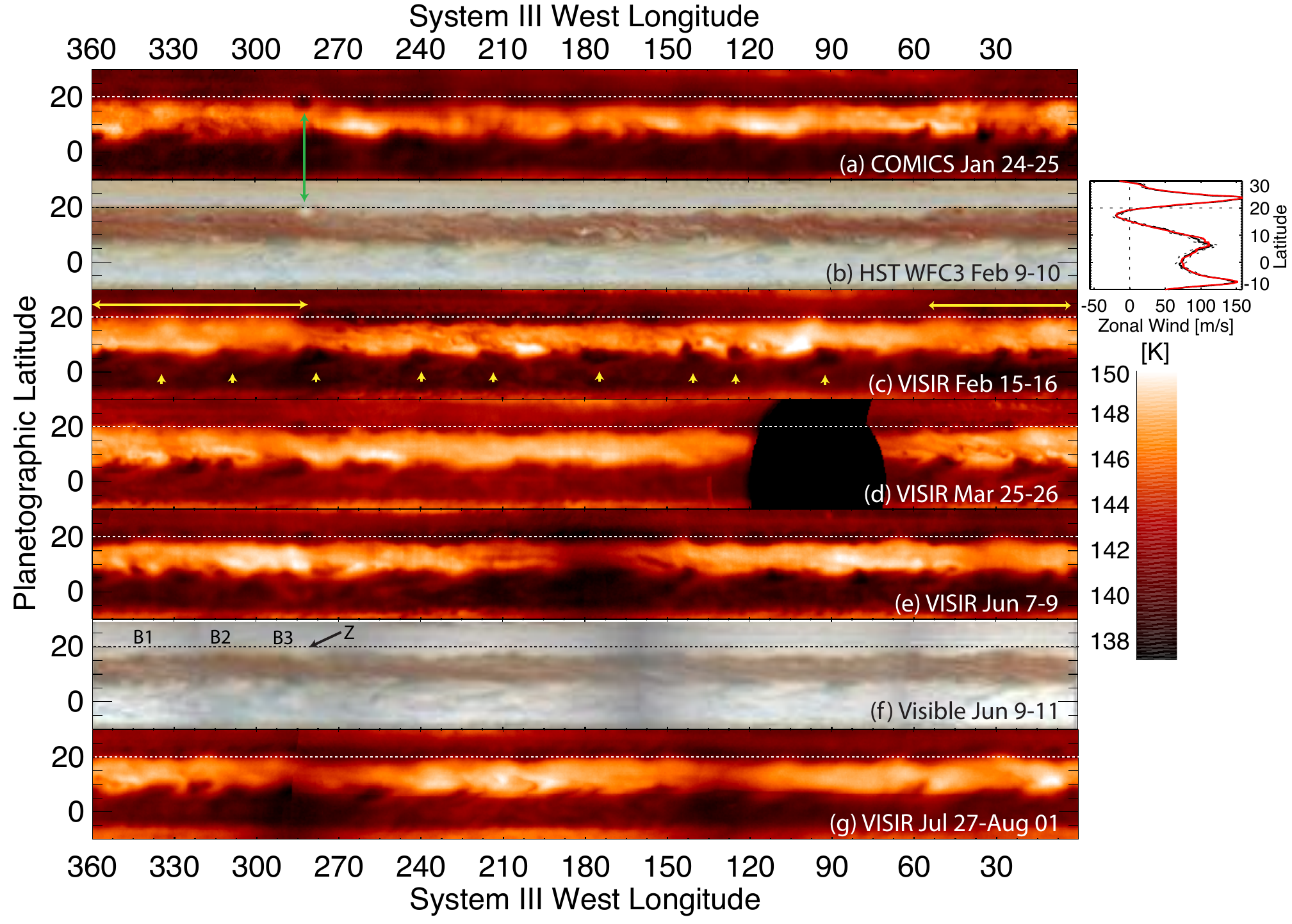}
\caption{Time series of 8.6-$\mu$m brightness temperature maps, January-August 2016.  Yellow arrows denote the expanded sector; barges B1-B3 are labelled in panel f.  Green arrows in panel b show White Spot Z in the west and two small red ovals in the east.  The visible-light maps in (b) were provided by Hubble WFC3 in February (see Fig. \ref{figS5}), and by amateur observers in (f) in June.  Hotspots and plumes (denoted by small yellow arrows) can be seen as bright and dark features on the southern edge of the NEB, respectively (Supplemental Text S4).  Horizontal lines at $20^\circ$N in each panel show the maximum extent of the expanded sector.  Cloud-top zonal winds from Hubble imaging in February 2016 (black) and January 2015 (red) are shown next to (b), indicating no change in either the jet location or speed \citep{17tollefson}. }
\label{expansion}
\end{figure}

The evolution of the 2015-16 NEB expansion event has been characterized in both visible-light observations from amateur observers (Supplemental Text S3, Figs. \ref{figS3}-\ref{figS4}) and in the thermal-infrared (Fig. \ref{expansion}).  This can be compared to Hubble WFC3 imaging in February 2016 (see also Fig. \ref{figS5}).  The exact starting date of the expansion is uncertain, as no identifiable plumes are involved.  Bulges of dark NEB colors gradually appeared near $16-18^\circ$N, west of the prominent White Spot Z (WSZ) near $19^\circ$N, $283^\circ$W, which has persisted since 1997 \citep{13rogers_rep3}.  These dark bulges were present as early as October 2014 (Text S3) and are visible in Hubble WFC3 imaging in January 2015 \citep[Fig. \ref{figS5};][]{15simon}, but appeared ephemerally between March-June 2015 (Fig. \ref{figS3}).  By October 2015, the expanded sector spanned $80^\circ$ of longitude between $320^\circ$ and $40^\circ$W.  This sector expanded east, reaching the WSZ, and west to a dark cyclone-anticyclone pair near $50-60^\circ$W, through January 2016 by infilling gaps between NEB bulges, occupying $\sim120^\circ$ of longitude at the start of 2016 (Fig. \ref{expansion}).    

Figure \ref{expansion} shows that the northern edge of the expanded NEB was at a higher latitude ($20^\circ$N) than elsewhere ($17^\circ$N), having replaced the white appearance of the NTrZ with darker colors.  The NEB appears bright at 8.6 $\mu$m due to a combination of warmer temperatures and low aerosol opacity, and the expanded sector of the NEB appears notably brighter than the undisturbed NTrZ ($50-270^\circ$W), consistent with the enhanced 5-$\mu$m brightness of the 2009-10 expansion event \citep{17fletcher_seb}.  

Previous NEB expansions have encircled the entire globe (Text S1), but the 2016 expansion only reached a maximum of $\sim145^\circ$ longitude by March and then regressed from both ends,  re-revealing the NEB dark bulges.  VISIR images in March 2016 suggested that the northern edge of the expanded NEB ($17-20^\circ$N) was dimming at 8.6 $\mu$m, re-establishing aerosol opacity over the NTrZ.  Visible-light images (Fig. \ref{figS4}) show a continued regression in April and May, and by June (Fig. \ref{expansion}e-f) the expanded sector had returned to normal, although some faint orange coloration could still be seen in the previously-expanded sectors.   The 8.6-$\mu$m-dark NTrZ could be seen to extend around the planet once more.  Thus, neither visible-light images from JunoCAM \citep{17orton_juno} nor 5-$\mu$m images from Juno/JIRAM showed any signs of the expansion at the first Juno perijove on 27 August 2016.  The exception is the presence of four newly-formed dark brown `barges' near $16-17^\circ$N, B1-B4 (Fig. \ref{expansion}f), that formed in the NEB during the expansion phase (Text S3).  After the recession of the expansion, these barges remained visible on the northern edge of the `normal' NEB.  


\subsection{Thermal changes during an NEB expansion}

\begin{figure}[h]
\centering
\includegraphics[width=18cm]{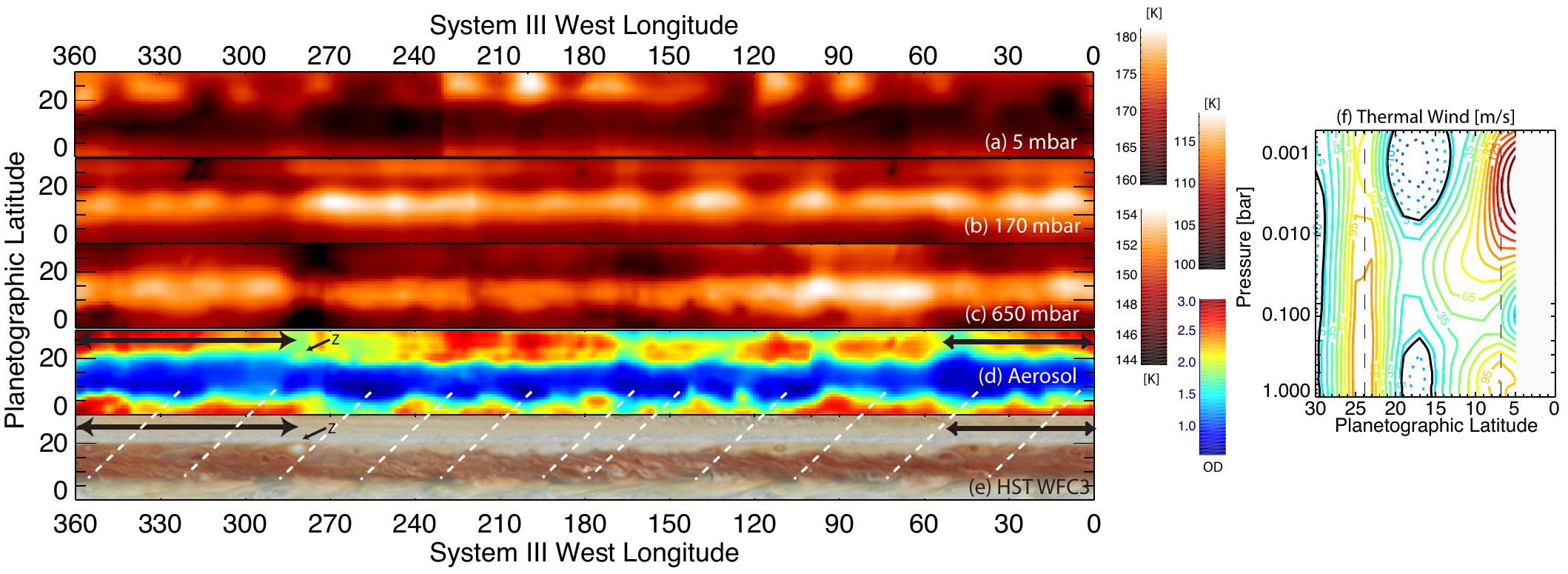}
\caption{Retrieved maps of temperature and aerosol opacity using VLT/VISIR observations on 15-16 February 2016.  Panels a-c show temperatures at 5, 170, and 650 mbar with uncertainties (Fig. \ref{figS7}) of $\pm5$, $\pm3$ and $\pm3$K, respectively.  Panel d shows the cumulative aerosol optical depth at 10 $\mu$m, integrated to the 1-bar level, assumed to comprise NH$_3$ ices in a $\sim600-800$ mbar cloud layer.  Only aerosol optical depth is varied, other aerosol properties (absorption cross section, single scattering albedo and vertical structure) are assumed to be invariant. Panel e shows the Hubble WFC3 map from 9-10 February (Fig. \ref{expansion}b), with white dashed lines showing the cloud-free hotspot locations in d-e.  The expanded sector of the NEB is indicated by black horizontal arrows in d-e.  Cold, cloudy plumes lie immediately east of the hotspots.  Zonal winds are shown in f, estimated from TEXES observations in December 2014 \citep{16fletcher_texes}.  Dashed vertical lines show the locations of the NTBs and NEBs eastward jets, solid black contours indicate $u=0$ m/s, and dashed contours indicate westward winds.}
\label{retrieval}
\end{figure}

What physical conditions are associated with the NEB expansion?  Either (i) a low-albedo brown aerosol forms over the NTrZ, or (ii) some dynamical process removes white aerosols from the NTrZ so that its coloration matches the NEB.  Hypothesis (ii) is akin to aerosol clearing events following a whitening of the South Equatorial Belt (SEB) \citep{11fletcher_fade}, and was suggested for the 2009-11 expansion event \citep{17fletcher_seb}.  Maps at eight wavelengths (7.9, 8.6, 10.7, 12.3, 13.0, 17.6, 18.7 and 19.5 $\mu$m) acquired over two nights (15-16 February) were combined to retrieve temperatures and aerosols in Fig \ref{retrieval}.  Absolute values are subject to large uncertainties resulting from the low information content of the 8-point spectra (see Fig. \ref{figS6}), but relative contrasts are robust \citep{09fletcher_imaging}.  Temperatures are shown at representative pressure levels in the stratosphere (5 mbar), upper troposphere (170 mbar) and near the cloud-tops (650 mbar). The vertical structure of the zonal winds is also shown \citep{16fletcher_texes}.  

The broad cool NTrZ at $19-24^\circ$N shows no evidence for thermal changes associated with the expanded NEB sector in the upper troposphere (100-500 mbar, Fig. \ref{retrieval}b). However, temperatures at 650 mbar (Fig. \ref{retrieval}c) suggest a 3-4 K warming in a diffuse warm area over the expanded sector at cloud top, coupled with a fall in the tropospheric aerosol opacity from $\sim2.8$ (representative of the non-expanded sectors) to $\sim1.5$ for the expanded sector (Fig. \ref{retrieval}d).  Whilst still not as aerosol-free as most of the NEB (opacities < 1.0), this reduction of opacity, coupled with the moderate rise in temperature, suggests evaporative removal of the NTrZ white aerosols to reveal deeper red-brown chromophores, supporting hypothesis (ii).  We caution the reader that temperature and aerosol changes are degenerate.  Nevertheless, Hubble WFC3 observations at 890 nm (Figs. \ref{tropwave}d \& \ref{figS5}), sensing upper-tropospheric aerosols, confirm a lower reflectivity over the expanded sector, consistent with the removal of white NTrZ aerosols.

This association between tropospheric warmth and aerosol clearing is similar to that observed during the revival of Jupiter's SEB in 2011 \citep{17fletcher_seb}, where large-scale subsiding airmasses removed aerosols in the regions surrounding convective plumes.  However, no such localized vigorous plumes were associated with the NEB expansion.  Vertical motions causing the NTrZ aerosol clearing could be related to the wave causing the ephemeral bulges on the northern edge of the NEB, but high-resolution wind measurements of any jet meanderings are not available (particularly regarding interaction between WSZ and the NEBn jet).  Zonal winds derived from Hubble observations in January 2015 and February 2016 (Fig. \ref{expansion}b) show no evidence for changes associated with the expansion \citep{17tollefson}.

\subsection{Upper tropospheric NEB wave}
\label{wave2}

Although the NEB expansion had regressed completely by Juno's arrival, thermal waves identified in the 2016 data were still active.  Fig. \ref{retrieval}b shows a wave pattern in the mid-NEB in the upper troposphere ($p<500$ mbar), similar to that observed by previous studies \citep[e.g.,][]{90magalhaes, 94orton, 97deming, 06li, 16fisher}.  The warm crests have a central latitude of $13-15^\circ$N, and the wave pattern is latitudinally confined to the bright belt at $10-17^\circ$N ($\pm1^\circ$).  We measure temperature excursions of 1.5-4.0 K with a mean amplitude of 2.5 K at 170 mbar (Figs. \ref{retrieval}, \ref{figS7}-\ref{figS8}).  The wave dominates the appearance of the upper troposphere and extends around the whole planet with a wavenumber $N\approx16-17$ in January-February 2016, including non-expanded NEB sectors.  This equates to a horizontal wavelength of $25700-27400$ km, making the mid-NEB wave distinct from the plume/hotspot wave on the NEBs, which has $N\sim12$ during this period (Supplemental Text S4).  The mid-NEB wave has no obvious counterpart in the main cloud deck.  It is not detected in the stratosphere, consistent with observations in 2000 \citep{06li}.

\begin{figure}[H]
\centering
\includegraphics[width=14cm]{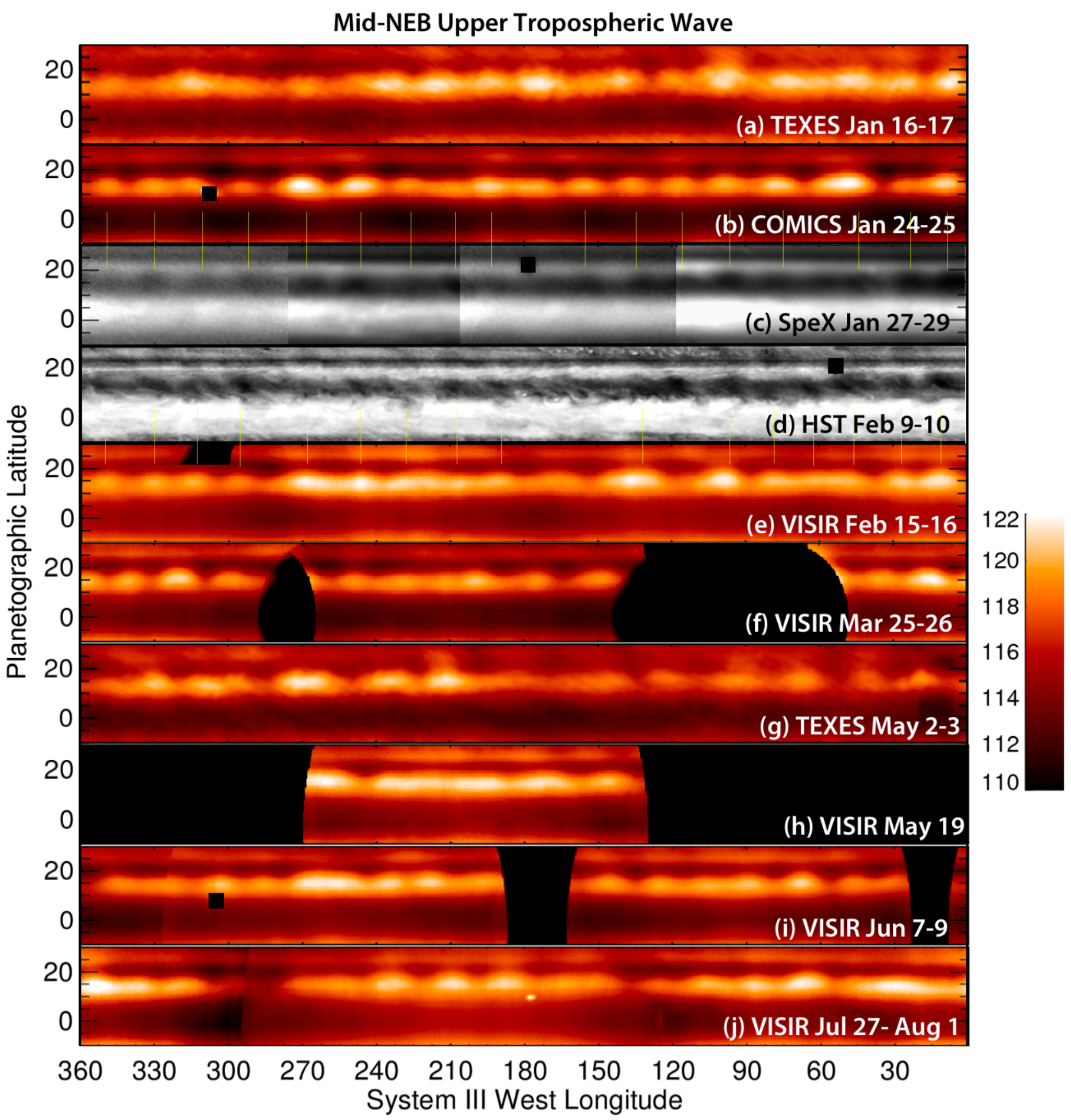}
\caption{Time series of 17.6-$\mu$m brightness temperatures (sensing $\sim150$-mbar temperatures) between January and August 2016. IRTF/TEXES maps are an average of the 584-589 cm$^{-1}$ spectrum and have a lower spatial resolution than the VLT/VISIR and Subaru/COMICS images. The wave is also shown in tropospheric aerosol reflectivity maps from IRTF/SPeX at 2.16 $\mu$m (c) and Hubble WFC3 at 890 nm (d), with vertical yellow ones showing their correspondence to the thermal wave (see Fig. \ref{figS9}).  Small black squares denote where Jovian satellites have been removed from panels b, c, d and i, resulting in small black squares.  Spatial coverage gaps are shown as black. }
\label{tropwave}
\end{figure}

Fig. \ref{tropwave} shows a time series of the tropospheric wave in 17.6-$\mu$m brightness temperatures (sensing $\sim150$ mbar) from Subaru, VLT and IRTF observations.  Contrasts in the wave pattern appear to vary, but this could simply result from different observing conditions on different nights.  We were unable to determine the phase speed of the wave from the thermal-IR data alone, as the gap between observations was too large, given the intrinsic variability of the wave pattern.  However, inspection of near-infrared reflectivity observations in strong CH$_4$ bands, 890 nm from Hubble WFC3 and 2.16-$\mu$m observations from the NASA/IRTF SpeX instrument (Fig. \ref{figS9}), taken between the COMICS observation on January 24-25 and the VISIR observation on February 15-16, reveal a very similar wave pattern in the upper tropospheric aerosols.  Fig. \ref{tropwave} demonstrates that mid-NEB temperatures are anti-correlated with aerosol reflectivity due to either (i) aerosol evaporation; or (ii) aerosol subsidence to deeper levels in the warm peaks of the mid-NEB wave.  Such a correlation was previously shown for the mid-NEB wave during Cassini's observations of an expansion event \citep{06li}.  Comparing the wave locations in these closely-spaced near-IR and thermal-IR maps reveal little evolution during this interval, suggesting that the wave pattern is either very slow or stationary, as confirmed by inspection of the individual 2.16-$\mu$m maps used for Fig. \ref{tropwave}c in Fig. \ref{figS10}, which shows no detectable motion during January 26th-29.  

A survey of previous detections of mid-NEB waves (Supplemental Text S2) reveals an observational bias to periods of previous NEB expansion events, so it is not known whether they are \textit{unique} to periods of expansion.  Reported wavenumbers have ranged from $N\sim5-7$ \citep{16fisher} to $N\sim12-13$ \citep{06li}, with 2-4 K amplitudes at 100 mbar and slow phase speeds $<5$ m/s.   The 2016 thermal wave ($N\sim16-17$) is therefore the most compact (highest wavenumber) yet observed.  Intriguingly, \citet{94orton} reported an increase in wave activity between 1988-90 and again in 1993 (compared to their 1978-1993 record), both coinciding with expansion events. The wave pattern characterized by Cassini \citep{06li} also occurred during an expansion.  Furthermore, comparison of 890-nm maps acquired by Hubble \citep{15simon} between 2015 and 2016 suggest that the reflectivity wave was not present (or had a lower amplitude) in 2015 (Fig. \ref{figS5}).  Nevertheless, without a comprehensive time series during non-expansion periods, any relationship between expansion events and the mid-NEB wave could simply be coincidental.

Besides the quasi-stationary nature of the 2016 wave and the anti-correlation between temperatures and haze reflectivity, the inversions in Fig. \ref{retrieval} also show that the wave is confined in both latitude and altitude (not present for $p>500$ mbar or at 5 mbar).  All of these characteristics are similar to those of the 2000-01 wave, so we follow \citet{06li} in identifying this as a Rossby wave.  Rossby-wave confinement occurs due to absorption, wave breaking and other wave-mean-flow interactions at critical surfaces, where the phase speed $c_x$ approaches the zonal velocity $u$ \citep{96achterberg}.  Furthermore, Rossby waves drift westward with respect to the zonal flow, so can only be close to stationary in eastward flow environments \citep{61charney}.  Zonal winds estimated from the thermal wind equation in December 2014 by \citet{16fletcher_texes} (Fig. \ref{retrieval}f) confirm that the NEB flow is positive (eastward) in the upper troposphere but negative (westward) at both the cloud-tops ($p>200$ mbar) and in the stratosphere ($p<7$ mbar), with $u\approx c_x \approx 0$ surfaces serving to confine the quasi-stationary wave to the upper troposphere.

Previous studies have suggested multiple origins for these NEB thermal waves, ranging from forcing by phenomena in the deeper interior \citep{90magalhaes, 97deming}; forcing by tropospheric meteorology and convective plumes \citep{87andrews}; flow disturbances around vortices \citep[e.g., around White Spot Z,][]{06li}; instabilities associated with the changing sign of the potential vorticity gradient \citep{96achterberg, 06read_jup, 16rogers_wave}; or the breaking of low-phase-speed Rossby waves generated by convection in the equatorial region, propagating poleward and depositing energy in regions where $u$ approaches zero \citep{16fisher}.  This latter mechanism was based on numerical simulations able to qualitatively reproduce observations \citep{16fisher}, and therefore does not require any flow perturbations related to an NEB expansion.  A comprehensive time series of wave variability is required to distinguish between these hypotheses.




\subsection{Stratospheric NTBs Wave}
\label{wave3}

Filters sensing Jupiter's CH$_4$ emission at 7.9 $\mu$m reveal stratospheric temperatures near the 5-mbar level.  The resulting temperature map (Fig. \ref{retrieval}a) indicates prominent wave activity over the $20-30^\circ$N region, covering the NTrZ and North Temperate Belt (NTB), centred on the strong eastward jet at $23.8^\circ$N (the NTBs).  The mean amplitude of the wave is $7.3\pm2.2$K at 5 mbar, but with some contrasts up to $\sim11$K (Fig. \ref{figS7}).

\begin{figure}[h]
\centering
\includegraphics[width=16cm]{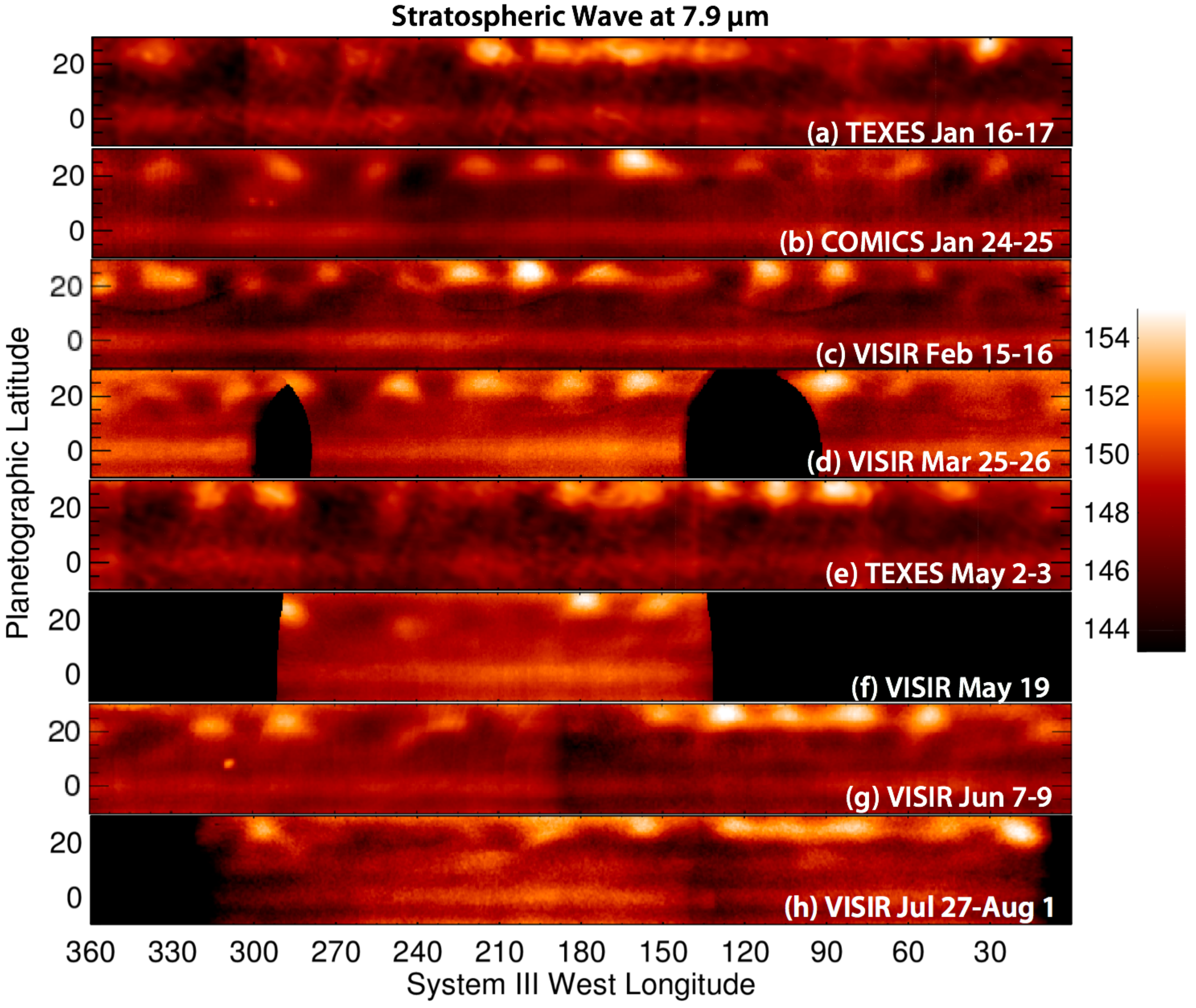}
\caption{Time series of 7.9-$\mu$m brightness temperatures (sensing $\sim5$-mbar temperatures) between January and August 2016. IRTF/TEXES maps are an average of the 1243-1252 cm$^{-1}$ spectrum and have a lower spatial resolution than the VLT/VISIR and Subaru/COMICS images.   Jovian satellites add artefacts to panels b and g near $300^\circ$W.  Spatial coverage gaps are shown as black.  }
\label{stratwave}
\end{figure}

Fig. \ref{stratwave} utilizes brightness temperature maps at 7.9 $\mu$m from IRTF, Subaru and VLT to trace the evolution of this wave over 6 months.  Unlike the tropospheric wave, the stratospheric wave exhibits more gaps in the wave train.  We observe $N\sim9-13$ temperature maxima (Fig. \ref{figS8}), depending on the visibility and the longitudinal coverage of each map.  Dates in early 2016 with the most complete longitudinal coverage yield horizontal wavelengths of 32000-38000 km at $24^\circ$N ($N=11-13$), making the stratospheric wave distinct from the mid-NEB wave.

Repeated observations in January (TEXES and COMICS) and May (TEXES and VISIR) suggest negligible motion of the wave pattern (Fig. \ref{stratwave}).  However, the motion is irregular and could simply represent the changing visibility of the wave crests with time.  Warm spots either side of a gap in the wave train from $\sim200-270^\circ$W suggest an eastward phase speed of $<0.1$ deg/day ($<5.5$ m/s) in May and June 2016, although we caution the reader that this may not be a unique interpretation.  Stratospheric zonal winds estimated from the thermal wind equation (Fig. \ref{retrieval}f) show that the velocity of the eastward NTBs jet does not vary much with altitude, so this wave has a westward phase speed with respect to the region of eastward $u$ from $20-29^\circ$N, consistent with a slowly-moving Rossby wave in the stratosphere.

Previous reports of stratospheric wave activity are sparse, given the difficulties of observing at 7.9 $\mu$m.  \citet{91orton} identified $N=11$ waves near $22^\circ$N from September to December 1988 (during an expansion), and a stratospheric wave ($N\approx11$) is evident over the $20-30^\circ$N region in the 1-4-mbar temperature maps from Cassini in 2000-01 \citep{04flasar_jup, 06li}.  IRTF/TEXES first identified stratospheric wave activity over a limited longitude range to the north and east of WSZ in December 2014, at a time when the northern edge of the NEB was undulating as a precursor to the 2015-16 expansion \citep{16fletcher_texes}.  A more comprehensive time-series of 7.9-$\mu$m observations, spanning several expansion events, is required to determine any relationship between these wave phenomena.

\section{Conclusions}

An Earth-based infrared campaign has characterized Jupiter's atmospheric variability at northern latitudes.  The poleward edge of the brown NEB appeared to broaden from $17^\circ$ to $20^\circ$N from November 2015 to March 2016, warming the atmosphere at $p>500$ mbar and removing white aerosols from the neighboring zone to reveal darker chromophores beneath.  This was the eighth such expansion event since 1987 (part of a 3-5-year cycle), but unlike previous events it did not progress to all longitudes, instead remaining limited to a longitude sector $\sim120^\circ$ west of prominent White Spot Z.  After March the event regressed southward, returning the NEB to normal conditions ahead of Juno's arrival, but leaving a chain of brown barges on the NEBn as evidence of the stalled expansion.  The 2016 observations also revealed striking wave patterns, distinct from the $N\approx12$ equatorial Rossby wave of plumes and hotspots on the NEBs jet ($6-9^\circ$N, Text S4).   Upper tropospheric temperatures in the mid-NEB ($13-15^\circ$N) exhibited a 2.5-K oscillation at 170 mbar that spanned all longitudes.  This wavenumber $N=16-17$ pattern appeared to be quasi-stationary, showed an anti-correlation between temperatures and aerosol reflectivity, and was confined to a region of eastward zonal flow in the upper troposphere so that it was not detected at 5 mbar or at the cloud tops ($p>500$ mbar).  Similar mid-NEB waves with lower wavenumbers existed during expansion events in 1988-89, 1993-94 \citep{94orton} and 2000-01 \citep{04rogers, 06li}, although a comprehensive survey of wave amplitudes over time is required to remove observational bias and to identify any links with NEB expansions.  Furthermore, significant stratospheric wave activity was detected at $20-30^\circ$N, with wavenumber $N=11-13$, a small phase speed within an eastward zonal flow, and a mean amplitude of $7.3\pm2.2$ K at 5 mbar.  Precursors for this stratospheric wave were identified in December 2014 \citep{16fletcher_texes}, and similar stratospheric wave activity was present in 1988 \citep{91orton} and 2001 \citep{06li}.  

Progress in understanding these wave patterns requires investigations over two very different timescales: (i) shortening the interval between thermal maps to confirm the small phase velocities of the waves; and (ii) investigation of wave occurrences and amplitudes in the historical record of IR observations and their relation to the cycles of NEB activity.  The first may be achieved during the Juno ground-based support campaign in the coming years.  The second is required to test any relationship between the NEB expansion cycle and the existence of these waves.   
\appendix
\section{Supplemental Appendices}
\subsection{Supplemental S1: History of NEB Expansions 1987-2012}

\begin{figure}[h]
\centering
\includegraphics[width=16cm]{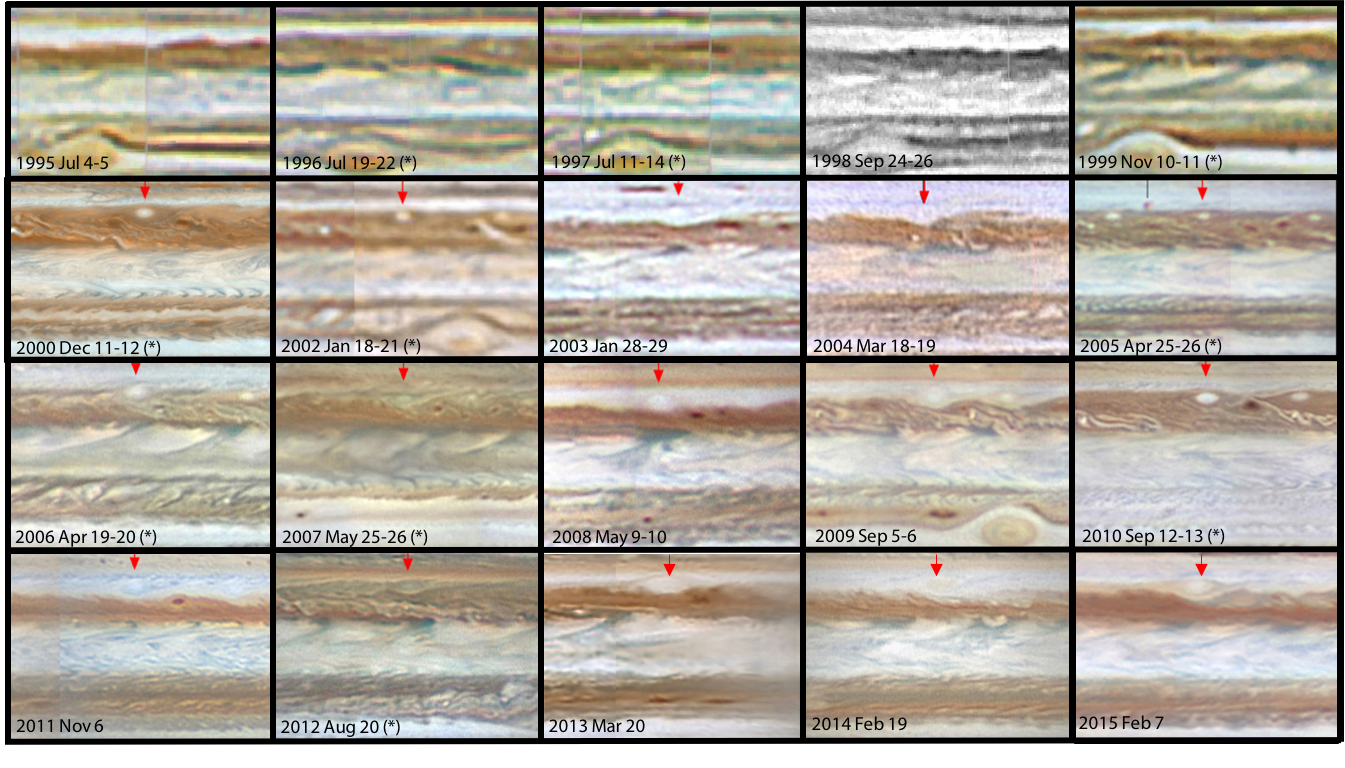}
\caption{Historical overview of NEB expansion events [3] through [7] (1995-2015), using images selected by \citep{17rogers}. Dates with an asterisk are during NEB expansion events as identified in Section S1.  Images are shown between ±30$^\circ$ planetographic latitude and were mapped in cylindrical coordinates (not equirectangular, as has been used elsewhere).   Images post-2000 are approximately centred on White Spot Z (red arrow, 19$^\circ$N).  The December 2000 image is from Cassini/ISS \citep{03porco} in equirectangular coordinates.  A full historical overview can be found in \citep{17rogers}.   A list of observers contributing to this figure can be found in Table S2.}
\label{figS1}
\end{figure}

This section provides an expanded description of previous NEB expansion events since 1987.  These events, which feature a darkening of the southern portion of the North Tropical Zone (NTrZ, $17-21^\circ$N latitude) that gives the appearance of the NEB expanding northwards, were first reported by \citet{95rogers}.  \citet{17rogers} published a major review of these events in the 1987-2010 time period, and we summarize the pertinent information here.  There were seven expansion events between 1987 and 2014.  The 1987-89 event was the first major event since the 1960s, and since then amateur observers have tracked expansion events with a 3-to-5 year period as detailed below and in Fig. \ref{figS1}.  Note that start times are easier to quantify than stop times, as the dark material becomes visible poleward of the $17^\circ$N NEBn (westward) jet, usually as a `bulge' on the northern edge of the NEB that is sometimes associated with a convective outbreak within the NEB.  The recession timescale is harder to quantify as it involves a gradual whitening (i.e., revival) of the white NTrZ aerosols. 

\begin{enumerate}
\item Expansion starting in December 1987, recession by early 1989, reaching a maximum latitude of $22.0^\circ$N in 1988.  This event was documented in \citet{95rogers}.
\item Expansion starting in March 1993, recession by mid-1994, reaching a maximum latitude of $20.3^\circ$N in 1994.
\item Expansion starting in April 1996, recession by November 1997, reaching a maximum latitude of $21.4^\circ$N in 1997.  This event can be seen by intercomparing Hubble WFPC2 images in October 1995 (no expansion) and October 1996 (during the expansion) in work by \citet{01simon} and \citet{01garcia}.
\item Gradual expansion throughout 1999 and 2000, recession by spring 2002, reaching maximum latitudes of $20.5^\circ$N in 2000-01 and $21.3^\circ$N in 2001-02. Cassini's December 2000 flyby of Jupiter imaged the planet during this epoch \citep{04rogers}.
\item Gradual expansion starting in April 2004, recession by April 2007, reaching a maximum latitude of $20.4^\circ$N in 2006.
\item Expansion starting in May 2009, recession by 2011, reaching a maximum latitude of $20.7^\circ$N in 2010.  This event was documented in both visible light and at 5-$\mu$m by \citet{11fletcher_fade, 17fletcher_seb}, demonstrating that aerosol clearing (i.e., enhanced 5-$\mu$m brightness) was associated with the darkened albedo of the NTrZ.  
\item Expansion starting after a convective outbreak in March 2012, preceded by a narrowing of the NEB through late 2011.  The expansion was complete by August 2012, and had regressed by early 2013 \citet{13rogers}.
\end{enumerate}

These events evolve in different ways and the progression is not always well organised - some start at a single longitude and spread around the planet, others start at several longitudes simultaneously in association with dark `bulges' on the northern NEB edge.  All have evolved such that the NEB had expanded at all longitudes, unlike the 2015-16 event described in the main article \citep{95rogers, 17rogers}.  \citet{17rogers} estimates 5-7 months for the completion of the expansion (except in 2000, where events proceeded more slowly), with a revival of the NTrZ white coloration (i.e., southward recession of the NEB) starting anywhere from 1-3 years post expansion, although these estimates are subjective and based on incomplete temporal sampling.  

\subsection{Supplemental S2: Previous Detections of Mid-NEB Waves}

In this section, we expand on previous detection and characterization of the mid-NEB waves (i.e., those detected in the upper troposphere spanning the width of the NEB) to provide context for the NEB wave described in the main article.  Note that this section does not deal with the trapped equatorial Rossby wave pattern on the NEBs \citep[and its associated hotspots and plumes,][]{90allison, 98ortiz, 00showman, 02baines, 05friedson, 06arregi, 13choi}.
  
Voyager IRIS infrared maps of Jupiter in March and July 1979 demonstrated significant longitudinal variability \citep{81hunt}.  \citet{89magalhaes} discovered a wavenumber-9 upper-tropospheric wave at $15^\circ$N that was further investigated by \citet{90magalhaes}.  They showed that this wave was indeed distinct from the NEBs wave, and that there was little displacement in the wave locations between the inbound and outbound maps.  This `fixation' in the System III longitude system led to the hypothesis that the wave was the signature of a deep-seated disturbance, although they cautioned that this was only one possible interpretation.
  
Further characterization of tropospheric thermal waves continued from ground-based observatories, particularly with the advent of 2D mid-infrared detectors. \citet{91orton} tracked stratospheric temperatures from 1980 to 1990 using 7.8-$\mu$m CH$_4$ emission; \citet{94orton} tracked 250-mbar temperatures from February 1978 to April 1993 using H$_2$ continuum emission at 18.2 $\mu$m.  They noted that the mid-NEB tropospheric wave was particularly strong between 1988-1990 (wavenumber $\sim10$ with a $\sim2$ K amplitude, with an estimated westward speed of 5.5 m/s), and again in 1993.  Furthermore, stratospheric waves near $22^\circ$N were also evident from September to December 1988.    These periods coincide with expansion events [1] and [2] in Section S1.  Observations in this period was supplemented by \citet{89deming, 97deming} between 1987 and 1993 during their search for planetary p-mode oscillations, identifying wavenumbers in the 2-15 range (with peaks at wavenumbers 6-11) and a westward phase speed of 5 m/s \citep{97deming}.  Unlike Orton's survey, the waves of Deming et al. appeared to cover the whole tropical domain.
  
Mid-IR imaging has continued since Galileo's arrival at Jupiter 1995, with a subset of results (June 1996 to November 1997) reported by \citet{16fisher}.  These identified mid-NEB waves of wavenumber 5-7, with a 2-3 K amplitude at 100 mbar and a low westward phase speed < 5 m/s.  This coincided with expansion [3] described in Section S1.  
  
With Cassini's December 2000 flyby, Jupiter was once again under scrutiny.  Images in methane absorption bands revealed a diffuse pattern with wavenumber  $\sim$14-18 over the NEB \citep{03porco, 04rogers}.   Spectral maps of Jupiter at 7-16 $\mu$m revealed a series of wave-like NEB features near 250 mbar \citep{04flasar_jup}, which were explored via near-simultaneous methane-band and thermal observations by \citet{06li}.  They revealed a wavenumber 12-13 wave spanning the NEB with a westward phase speed of 3.9 m/s.  They interpreted this as a Rossby wave, latitudinally confined by the curvature of the wind field, and confined in altitude by a region of eastward (prograde) flow in the 4-500 mbar range.  This mid-NEB wave was not observed in the stratosphere, although a stratospheric wave spanning 20-30$^\circ$N was active during this time.  This coincided with expansion [4] described in Section S1.
  
There are no further attempts to report NEB thermal wave phenomena in the literature since the Cassini flyby (although data do exist).  Ground-based spectroscopic maps in December 2014 from the TEXES instrument \citep{16fletcher_texes} do show evidence for a stratospheric wave, but the NEB structure was interpreted as being related to the NEBs Rossby wave (i.e., hotspots and plumes), rather than a mid-NEB wave.  Exploring the temporal variability of the NEB waves over the past three decades, particularly in relation to the NEB expansion events, will be the subject of future work.
  
\subsection{Supplemental S3: Chronology of the 2015-16 NEB Expansion}

\begin{figure}[h]
\centering
\includegraphics[width=16cm]{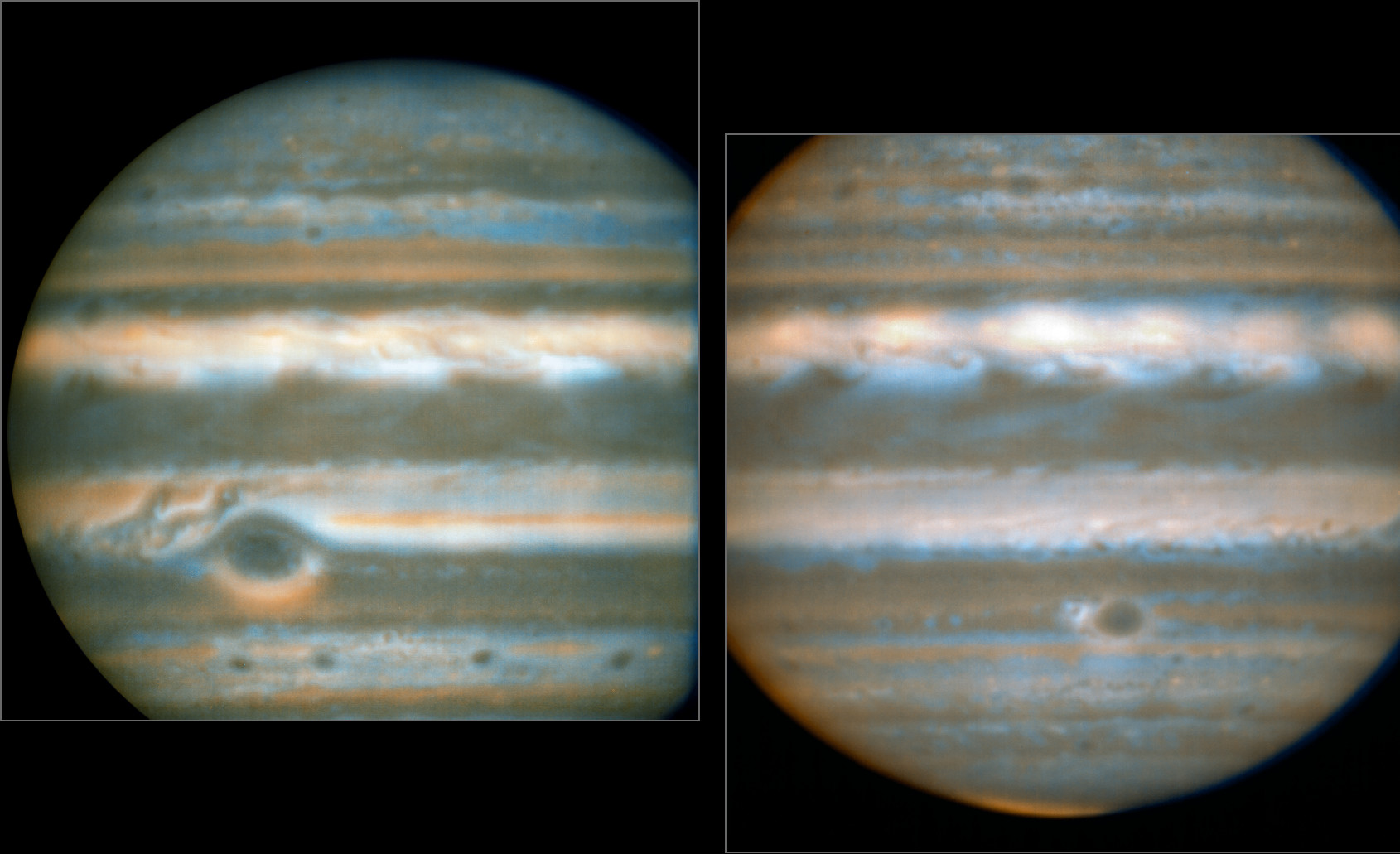}
\caption{Full-disc observations of Jupiter taken by VLT/VISIR on February 15th 2016 (left) and March 25th 2016 (right).  These images provide context for the 0-30$^\circ$N latitude range shown in the main article.  They are produced as a two-color composite, with blue=8.6 $\mu$m and orange=10.7 $\mu$m.  The mid-NEB wave is visible as the train of features in orange, sensing atmospheric temperatures in the 100-500 mbar range.  The cloud-free hotspots (blue, high 8.6-$\mu$m brightness) and associated broad, cloudy plumes (dark) are visible on the NEBs, just north of the equator.  The Great Red Spot (left) and Oval BA (right) are visible in the southern hemisphere, as is emission from ethylene in the south polar aurora (orange, right image).}
\label{figS2}
\end{figure}

\begin{figure}[h]
\centering
\includegraphics[width=12cm]{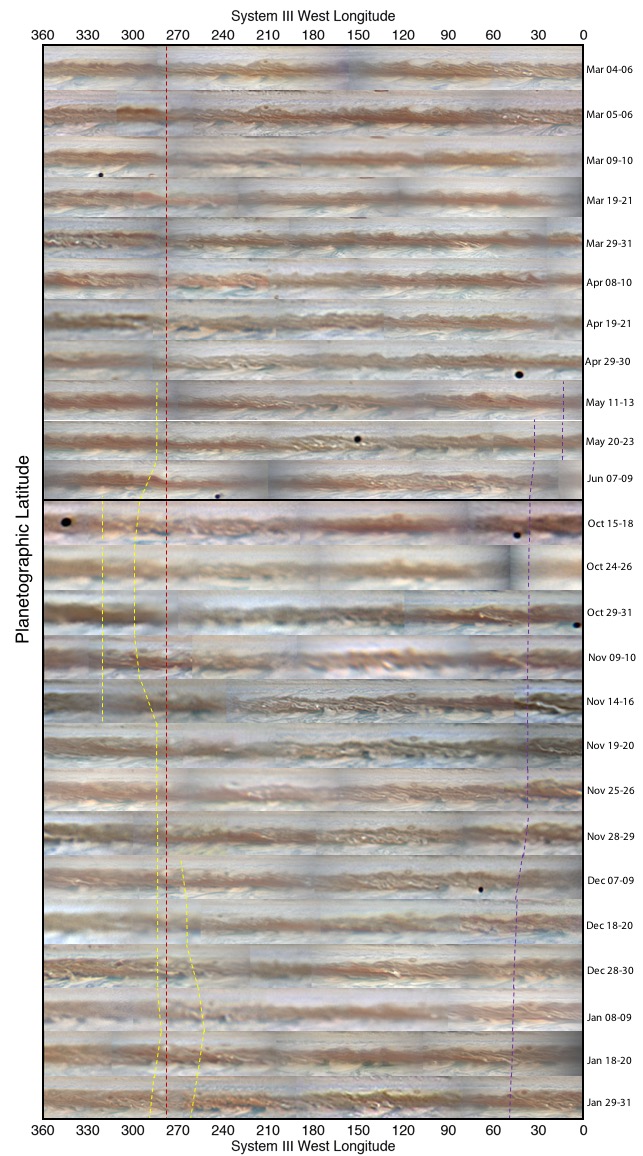}
\caption{Extracts from Jupiter maps assembled by the amateur astronomy community (credit: M. Vedovato) with a 7-10 day cadence, March 2015-January 2016.  Latitudes span 0-30$^\circ$N. Yellow/purple dotted lines attempt to track the eastern/western edge of the expanded sector, and there are multiple instances of these lines when there was ambiguity in the east-west extent of the expanded region.  Red dotted line shows the location of White Spot Z. }
\label{figS3}
\end{figure}

\begin{figure}[h]
\centering
\includegraphics[width=12cm]{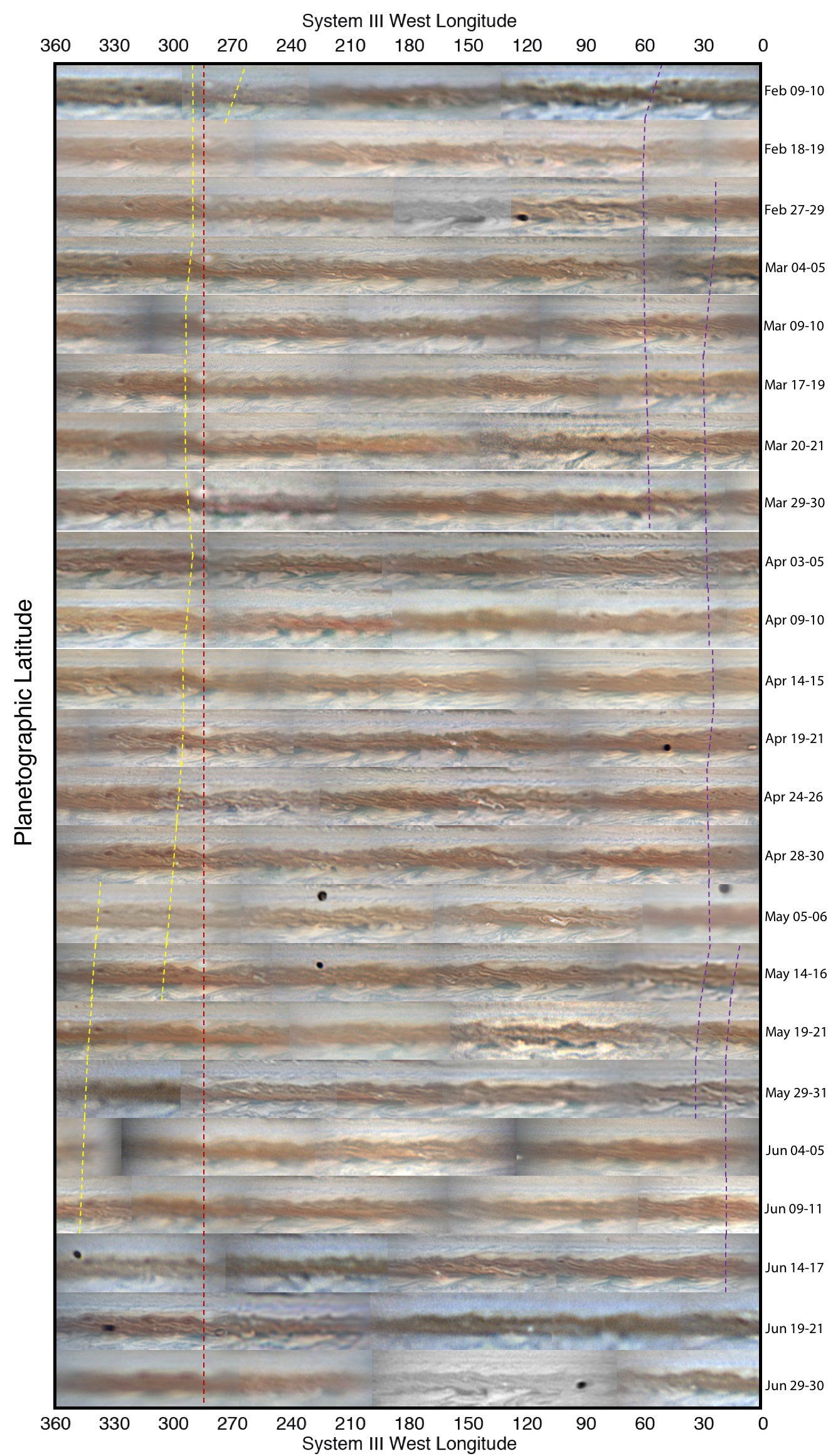}
\caption{As for Fig. \ref{figS3}, but covering February-June 2016. }
\label{figS4}
\end{figure}

Although this paper deals with the discoveries highlighted by the infrared imaging, we present here an expanded version of the chronology of the 2015-16 expansion event, as summarized in the main article.  Amateur observers regularly upload their images to servers such as PVOL and ALPO-Japan \citep[for full details of their observing workflow, see][]{14mousis_proam}.  M. Vedovato produces a database of global maps assembled from the amateur imaging using the JUPOS software (\mbox{www.jupos.org}), creating a map approximately every 7-10 days, and hosted on the Unione Astrofili Italiani website (\mbox{http://pianeti.uai.it/}).   We used extracts from these maps of the $0-30^\circ$N region to assemble the chronological charts in Figs. \ref{figS3} and \ref{figS4}, spanning March 2015 to June 2016.  

\begin{figure}[h]
\centering
\includegraphics[width=16cm]{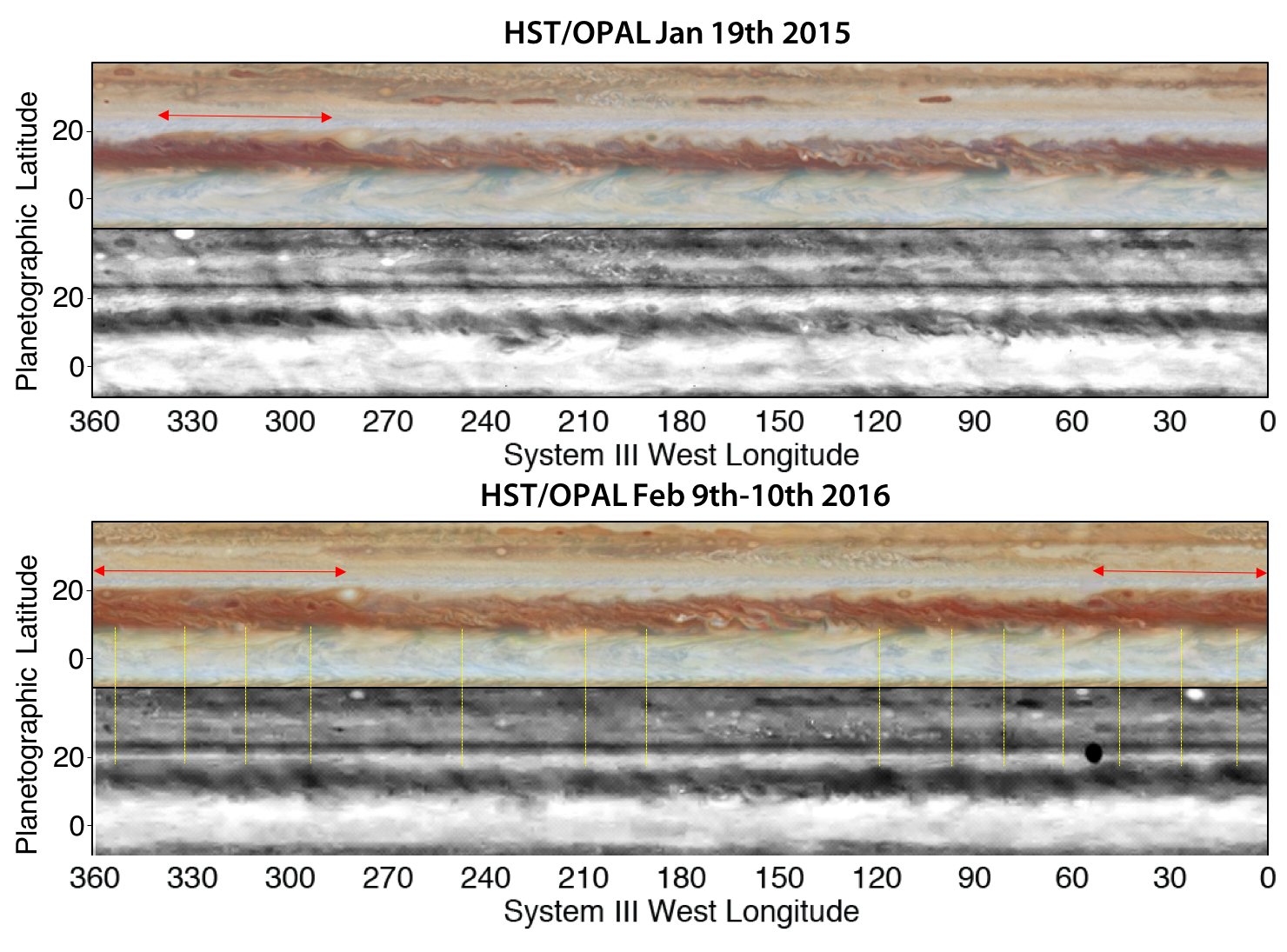}
\caption{Hubble Space Telescope WFC3 maps of Jupiter in 2015 and 2016 acquired as part of the OPAL program (Credit:  A. Simon, M. Wong, G. Orton), and available via the Mikulski Archive for Space Telescopes (\mbox{https://archive.stsci.edu/prepds/opal/}).  The top panel is a color map assembled from 397, 502 and 631nm filters.  The bottom panel is a CH$_4$ band image (889 nm) sensitive to high-altitude aerosols.  The expanded sector is shown by the red horizontal arrows (note that this is only faintly visible in the 889-nm map).  The vertical yellow lines in February 2016 show the troughs of a wave with 14 identifiable dark spots.  This wave appears to be most regular west of 285$^\circ$W, as far as 120$^\circ$W, and sometimes (but not always) aligns with bulges on the NEB northern edge.  The January 2015 image does not show comparable structure.  Diffuse diagonal striations are artefacts in the WFC3 maps most likely caused by long-wavelength fringe defects in 889-nm WFC3 images (\mbox{http://www.stsci.edu/hst/wfc3/documents/handbooks/currentIHB/wfc3\_ihb.pdf}). }
\label{figS5}
\end{figure}

These show the complexity and subjectivity in tracking an NEB expansion that has not progressed to completion, entirely spanning a longitude circle.  The northern edge of the NEB regularly features undulations of the dark NEB material and white NTrZ material, which we refer to as `bulges'.  Some of these are associated with dark, cyclonic circulations (`barges,' readily visible in Hubble imaging in January 2015, \citet{15simon} and Fig. \ref{figS5}) near $16-18^\circ$N, which form `bulges' along the edge of the NEB.  The NEB appears to expand when the dark material is visible in between the bulges (i.e., replacing the white NTrZ aerosols), such that the bulges are no longer independently visible, although some darker brown barges remain evident.  The lines in Figs. \ref{figS3} and \ref{figS4} are an attempt to delineate the easternmost (yellow lines) and westernmost (purple lines) edges of the expanded sector based on this criterion.  Where a gap opens up between two previously indistinguishable NEB bulges, a new line is added to the figure representing the growth or regression of the expanded sector.  However, such gaps often appear orange in color (i.e., intermediate between the white NTrZ and brown NEB), requiring the users' distinction between 'expanded' and 'non-expanded' sectors.

\citet{16rogers} reported preliminary signs of a disturbance on the northern edge of the NEB in October-November 2014, in a sector west of White Spot Z.  This can be seen as a faint orange sector between $280-340^\circ$W in Hubble OPAL imaging in January 2015 \citep{15simon} (Fig. \ref{figS5}), although the amateur time series in Fig. \ref{figS3} indicated that this was transient.  Further onsets of a broadening event may have occurred in May-June 2015, spanning some $\sim80^\circ$ of longitude west of White Spot Z, but pale gaps of NTrZ aerosols can still be seen on some of these dates.  After solar conjunction in August 2015, the expanded sector was more clearly defined between $320^\circ$W and $40^\circ$W (covering $\sim80^\circ$ of longitude in October 2015).   

Fig. \ref{figS4} shows the progression from October 2015 to January 2016, overlapping the first thermal-IR maps in the main article.  The western edge (purple line) moved from 40$^\circ$W to 50$^\circ$W during this time, appearing to meet (and move west of) a small dark-brown spot (barge).  This small dark spot is moving slowly westward during this interval.   The eastern edge moved east from 320$^\circ$W towards White Spot Z (283$^\circ$W) throughout November 2015, by filling in the pale gaps between the NEB bulges.  The expanded sector therefore covered approximately 120$^\circ$ longitude at the start of 2016.  There is also a subjective impression that the expansion extended further east of White Spot Z to 255$^\circ$W by late December/early January (so that this sector covered $\sim$150$^\circ$ longitude), although this brown extension east of White Spot Z faded and was not longer evident after the beginning of February 2016.

Fig. \ref{figS4} shows the stagnation and regression of this expanded sector from February to June 2016.  Between February and March the expanded sector was bordered by White Spot Z at its eastern end (283$^\circ$W) and by a cyclone-anticyclone pair near 50$^\circ$W.  Imaging by the Hubble OPAL program (February 9th-10th 2016) captured the expansion during this stage, Fig. \ref{figS5}.  In late March and April 2016, the western edge began to regress, as the expanded sector became orange in color between two bulges - this is shown by our choice of moving the western edge of the expansion closer to 30$^\circ$W.  At the same time, paler aerosols can be seen moving gradually west of White Spot Z, so that the bulges on the NEB edge were becoming visible (i.e., it was no longer a fully-expanded sector).  By mid-May, the 17-20$^\circ$N latitude range had changed from brown (maximum expansion) back to pale orange over the whole expanded sector, and this fading continued through May and June.  Fig. \ref{figS3}-\ref{figS4} demonstrate that the peak of the expansion occurred between November 2015 and April 2016, with regression thereafter.  

Brown barges (cyclonic circulations at 16$^\circ$N) are the remnants of the NEB expansion phase.  They are visible on the northern edge of the `normal' NEB at the end of the sequence.  Figs. \ref{figS3}-\ref{figS4} shows the clearest of these features was present from January 2016 (and possibly earlier), moving from near 320$^\circ$W in February 2016 to 345$^\circ$W by June 2016, consistent with the retrograde direction of the NEBn jet. The formation of these barges, within the expanded regions, is typical for each of the expansions studied by \citet{17rogers}.

\subsection{Supplemental S4: The NEBs Wave}

\begin{figure}[h]
\centering
\includegraphics[width=16cm]{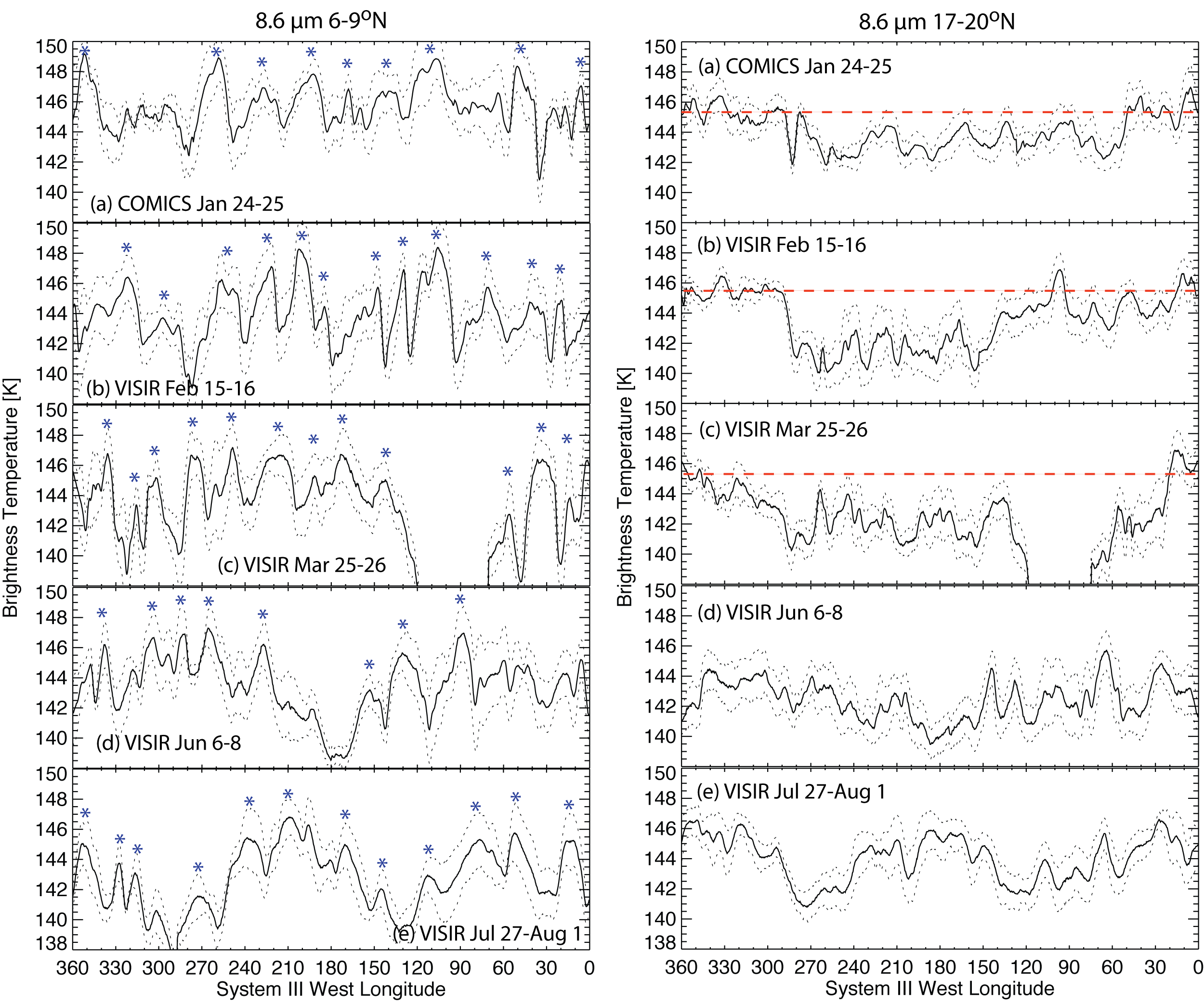}
\caption{East-west profiles through 8.6-$\mu$m COMICS and VISIR filters on five selected dates, extracted from maps shown in Fig. 1 of the main paper.  Panels on the left show cross-sections through the NEBs hotspot/plume wave between 6-9$^\circ$N, with the location of hotspots (cloud-free and bright at 8.6 $\mu$m) identified by blue stars.  Panels on the right show cross-sections at the northern edge of the NEB and southern edge of the NTrZ at 17-20$^\circ$N.  The horizontal red dashed line provides guidance for the contrast between the expanded sector (warm) and the non-expanded sector (cool) in January-March 2016.  Gaps in the longitude coverage exist in March 2016.}
\label{figS6}
\end{figure}

\begin{figure}[h]
\centering
\includegraphics[width=16cm]{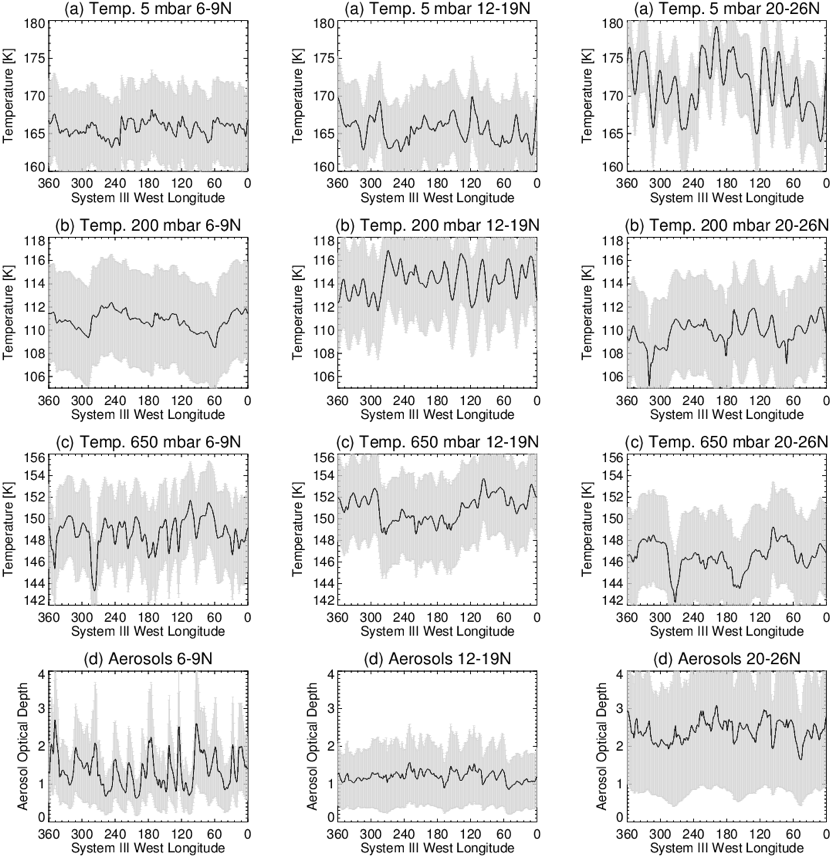}
\caption{East-west profiles of temperatures and aerosols retrieved from VLT/VISIR maps on February 15-16th 2016, as shown in Fig. 2 of the main paper.  These are cross-sections, averaged over the 6-9, 12-19 and 20-26$^\circ$N ranges pertinent to the NEBs hotspot-plume wave, the mid-NEB wave, and the stratospheric wave above the NTBs jet.   Temperatures are shown at 5, 200 and 650 mbar, whereas aerosols are represented as the cumulative 10-$\mu$m optical depth to the 1-bar level.  Grey error bars show the formal uncertainty in the retrieval process \citep[primarily due to poor vertical information content in retrievals from photometric imaging][]{09fletcher_imaging}.  Point-to-point precision of the measurements is much smaller than this.  The NEBs wave has the strongest signature in the aerosol retrievals for 6-9$^\circ$N; the mid-NEB wave can be seen at 200 mbar for 12-19$^\circ$N, and the stratospheric wave can be seen at 5 mbar for 20-26$^\circ$N.  Wave amplitudes and wavelengths are reported in the main article.}
\label{figS7}
\end{figure}

\begin{figure}[h]
\centering
\includegraphics[width=16cm]{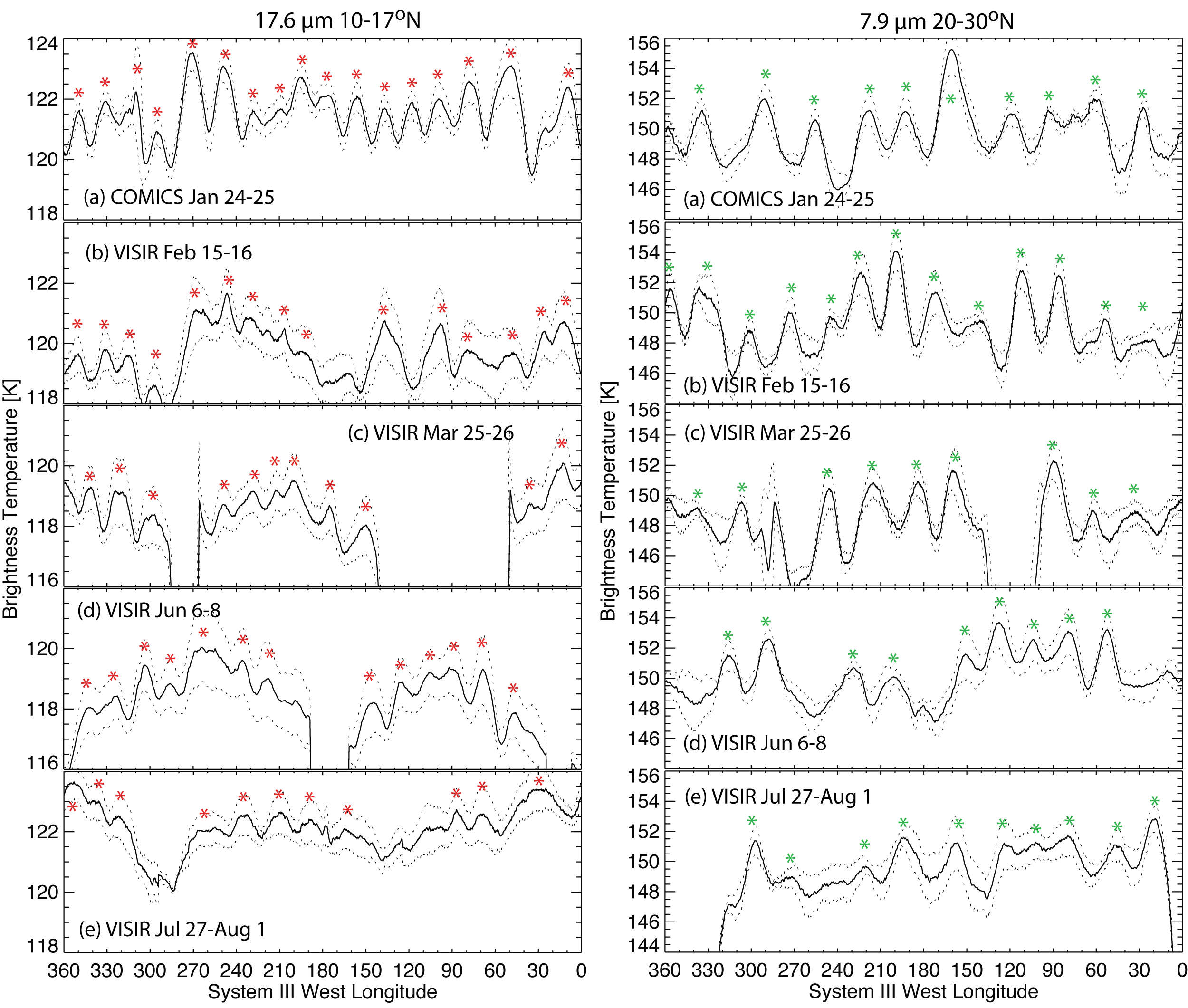}
\caption{East-west profiles through COMICS and VISIR filters on five selected dates.  The left panels show 10-17$^\circ$N cross-sections through the 17.6 µm brightness temperature maps from Fig. 3.  The identifiable peaks of the mid-NEB upper tropospheric wave are shown by red stars.  Limb darkening has been partially corrected in the 17.6-$\mu$m cross-sections, but is still evident as gradients over ~90$^\circ$ longitude scales. The right panels show 20-30$^\circ$N cross-sections through the 7.9-$\mu$m brightness temperature maps from Fig. 4. The stratospheric wave peaks are identified by green stars.  Error ranges represent the standard deviation of the latitudinal brightness temperature within each longitude bin.  Gaps in the longitude coverage are present for all dates after February 2016. }
\label{figS8}
\end{figure}

The main article covers two newly-identified wave patterns in the 0-30$^\circ$N domain, but the data presented throughout the article also shows thermal contrasts associated with a third wave on the prograde NEBs jet between 6-9$^\circ$N.  This pattern of visibly-dark, longitudinally-elongated `hotspots' and adjacent bright fans of material (referred to as plumes, although these are not always associated with convection) that can be seen near $8^\circ$N in reflected sunlight \citep{90allison, 02baines, 06arregi, 13choi}, infrared \citep{98ortiz, 16fletcher_texes}, and radio-wave observations \citep{16depater, 17cosentino}.  This chain moves slowly westward with respect to the rapid eastward flow of the NEBs, and has been interpreted as an equatorial Rossby wave pattern \citep{98showman, 00showman, 05friedson}, with rising air and condensation in the plumes and descent and aerosol-clearing in the hotspots. 

The maps in Fig. \ref{expansion} show the hotspots as a wavenumber $N\approx12$ pattern of visibly-dark, 8.6-$\mu$m bright features at $8.4^\circ$N (typically 8-12 individual features are seen).  Ammonia- and aerosol-rich plumes appear as dark fans of material (low radiance) at $6.4^\circ$N, to the southeast of the hotspots, and extending into the equatorial zone in a southwesterly direction, consistent with the latitudinal shear between the equator and the NEBs.  The cold plumes are visible at all wavelengths from 8-13 $\mu$m (sensing $p>400$ mbar), consistent with upwelling and adiabatic expansion.  They are not visible at 17-20 $\mu$m (sensing $p<400$ mbar), suggesting that they do not influence temperatures in the upper troposphere.  This is confirmed by the temperatures retrieved in Fig. \ref{retrieval}, where the plumes cannot be seen at 170 mbar, but are visible at 500-650 mbar.  

The 650-mbar temperature map (Fig. \ref{retrieval}d and Fig. \ref{figS7}) suggests that a 3-4 K contrast exists between the cold plumes (which can be readily seen in multiple N-band filters) and the hotspots (which can only be seen at 8.6 $\mu$m) in the $7-9^\circ$N latitude range.  The hotspots have aerosol optical depths as low as 0.7-1.0, compared to 2.5-3.0 for the plumes.  However, we caution that there is a significant degeneracy between temperature and aerosol opacity retrievals, such that the hotspot visibility at 8.6 $\mu$m could be the result of cloud clearing alone.  The upwelling and subsidence suggested by the temperature and aerosol retrievals support the identification of this pattern as a manifestation of a Rossby wave on the NEBs, extending from the deep atmosphere to the $\sim500$ mbar level.  This wave is distinct from both the mid-NEB wave and the stratospheric described in the main article.

\begin{figure}[h]
\centering
\includegraphics[width=16cm]{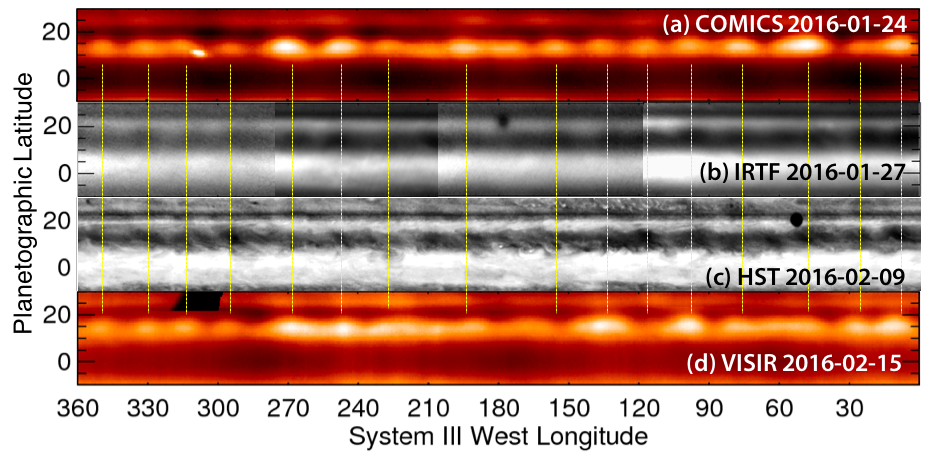}
\caption{Comparison of thermal-IR maps at 17.6 $\mu$m to reflectivity maps in bands of strong CH$_4$ absorption in January-February 2016.  The wave has low contrast in the CH$_4$-band imaging, even when we stretch the contrast.  Nevertheless, the yellow dashed lines indicate that the same wave train is visible in both thermal and reflectivity imaging.  Not all features remain identical over the 22 days between the COMICS and VISIR observations, but this figure supports the suggestion that the wave pattern is quasi-stationary in System III.   IRTF/SpeX observations at 2.16 $\mu$m (panel b) were taken over three consecutive nights (January 26th-28th 2016) and can be downloaded from \mbox{http://junoirtf.space.swri.edu}.     Hubble observations at 889 nm (panel c) were obtained on February 9th 2016 and are available from \mbox{https://archive.stsci.edu/missions/hlsp/opal/cycle23/jupiter/}.  A more comprehensive analysis of the wave train in CH4-band imaging will be the subject of a future publication, but this confirms an anticorrelation between warm temperatures and reduced aerosol opacity (and therefore reflectivity), as was found during the Cassini observations by \citet{06li}.}
\label{figS9}
\end{figure}

\begin{figure}[h]
\centering
\includegraphics[width=16cm]{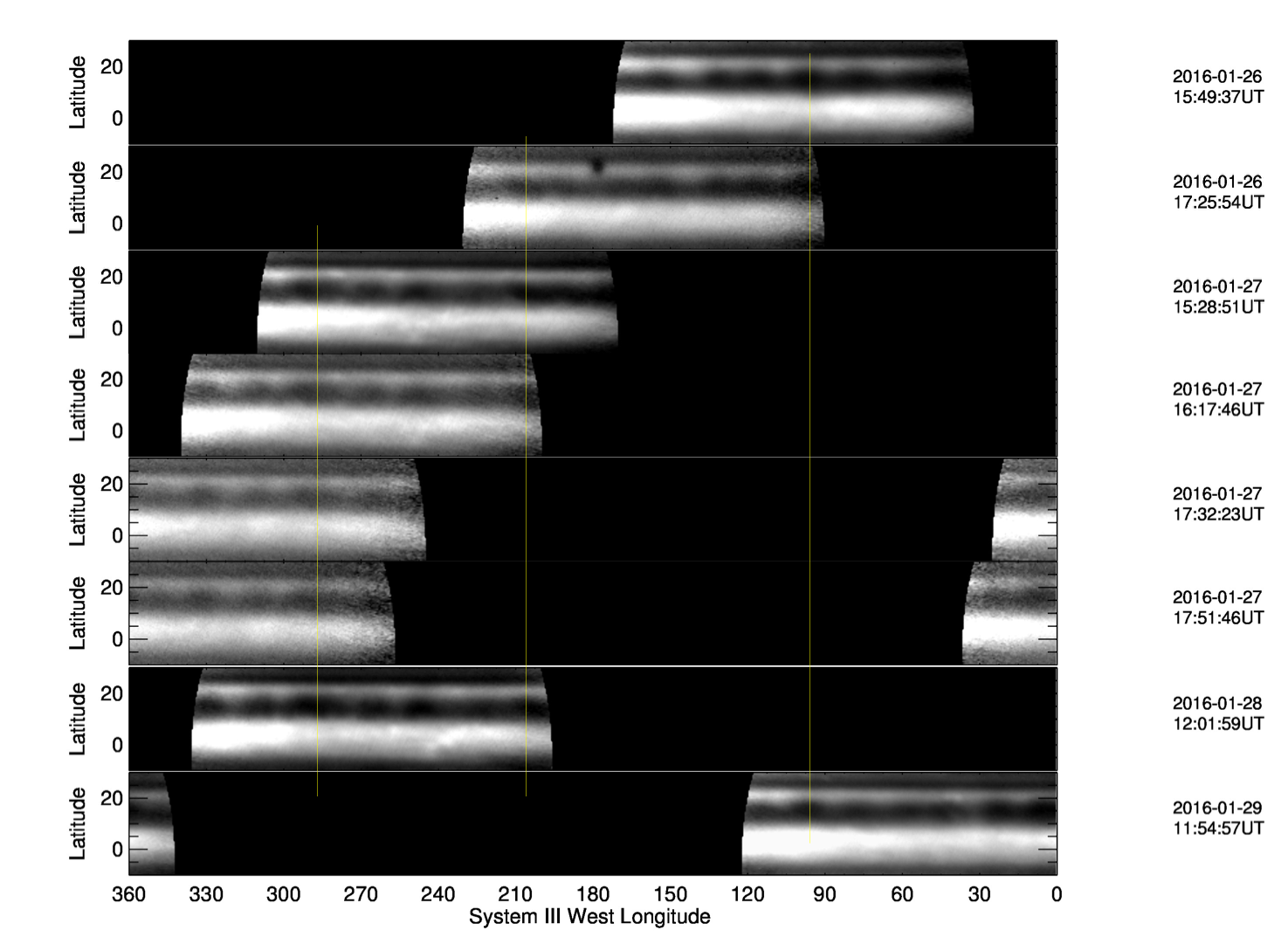}
\caption{Tracking of the mid-NEB wave at 2.16 $\mu$m in January 2016.  A combination of these individual maps is used in Fig. 3c of the main article.  Vertical yellow lines are included as a guide to show the location of similar parts of the wave train over multiple nights, supporting the suggestion that these are quasi-stationary features in System III.   These images were taken over four consecutive nights (January 26th-28th 2016) and can be downloaded from \mbox{http://junoirtf.space.swri.edu}.  They have been cylindrically mapped and corrected for limb darkening.}
\label{figS10}
\end{figure}

\begin{figure}[h]
\centering
\includegraphics[width=14cm]{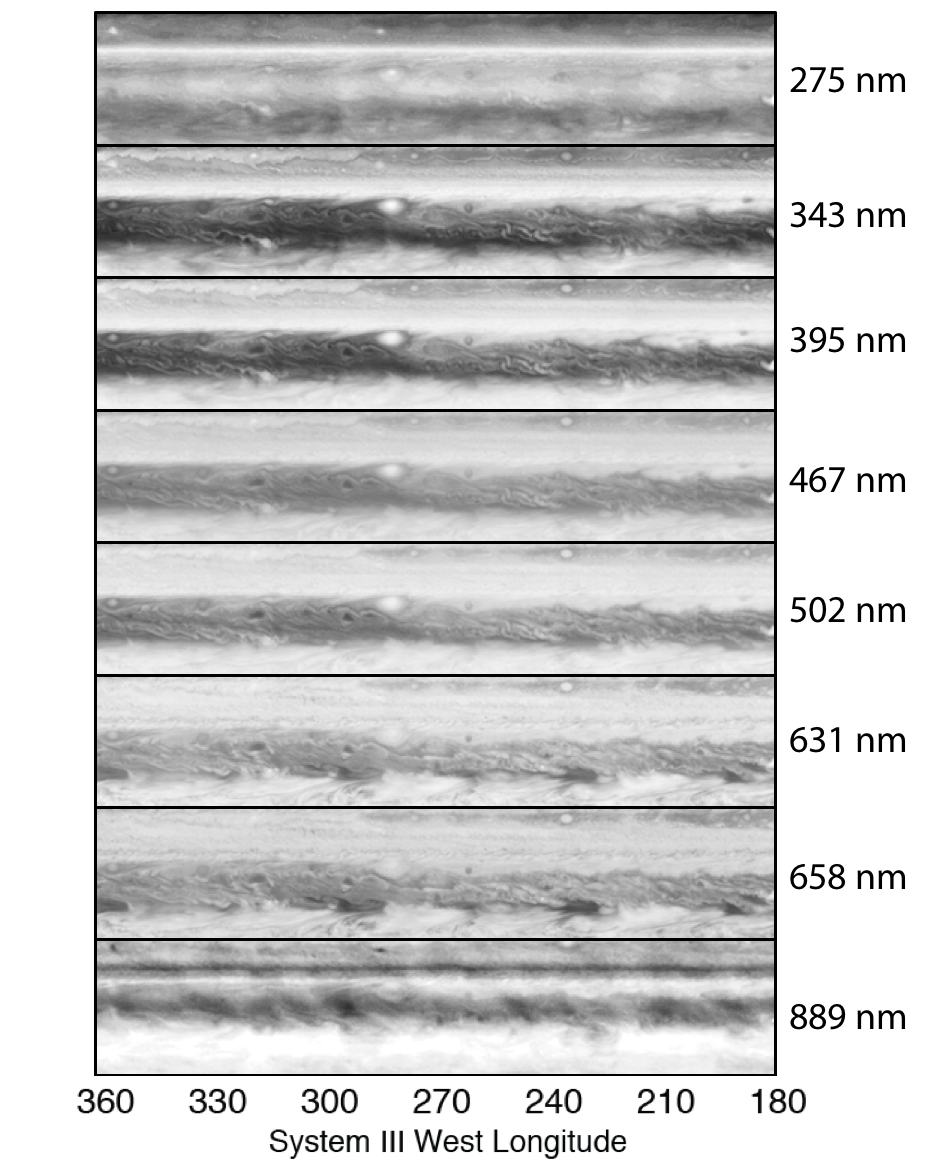}
\caption{Hubble WFC3/UVIS maps of Jupiter taken on February 9th-10th 2016.  Only the 180-360$^\circ$W longitude range is shown for 0-35$^\circ$N, highlighting contrasts east (non-expanded) and west (expanded) of White Spot Z at 283$^\circ$W.  The contrast is greatest at blue wavelengths (343-395 nm), consistent with the enhanced red coloration of the NTrZ west of White Spot Z.  This contrast diminishes through to red wavelengths (631-658 nm), but can still be observed.  In the methane absorption band at 889 nm, the expanded sector exhibits slightly lower reflectance from high-level aerosols than the non-expanded sector consistent with thermal infrared observations suggesting a partial clearing of aerosol opacity there.  The UV (275 nm) filter approximates to a negative of the 889-nm filter, because methane-bright high-level haze is generally dark due to UV absorption. Thus at 275 nm, the NEB is bright and the methane-dark waves appear especially bright.  However, the strip at $\sim$20$^\circ$N appears slightly darker west of White Spot Z than east of it, suggesting an anomalous haze layering in this expanded sector.  All observations were obtained from: \mbox{https://archive.stsci.edu/missions/hlsp/opal/cycle23/jupiter/}. }
\label{figS11}
\end{figure}

 \begin{table}
 \caption{Observational details for the thermal-infrared observing campaigns in 2016.  All data is available from the primary author on request, and is archived on the ESO archive (\mbox{http://archive.eso.org/eso/eso\_archive\_main.html}) and Subaru archive (\mbox{http://smoka.nao.ac.jp/}). }
 \centering
 \begin{tabular}{cccc}
 \hline
  Telescope/Instruments & Dates & Wavelengths & Comments \\   
  \hline
NASA-IRTF TEXES 			& Jan 16-17 	& 17.1 and 8.0 $\mu$m 	& Spectral maps 584-589 cm$^{-1}$ \\
IDs 2015B018 and 2016A027 	& May 2-3 	& 				& (R$\sim$5800) and 1243-1252 cm$^{-1}$ (R$\sim$12400) \\
\hline
VLT VISIR						& Feb 15-16	& 7.9, 8.6, 10.7, 		& Global images over 2 nights \\
IDs 096.C-0091 and 097.C-0222	& Mar 25-26	& 12.3, 13.0, 17.7, 		& to create full maps. \\
							& May 19		& 18.6, 19.5 $\mu$m		& \\
							& Jun 6-7		&					& \\ 
							& Jul 27-Aug 1	&					& \\
\hline
Subaru COMICS			& Jan 24-25 & 7.9, 8.7, 17.6 $\mu$m & Global images over 2 nights \\
ID S16B-025 & & & to create full maps \\
 \hline
 \end{tabular}
 \end{table}

 \begin{table}
 \caption{List of amateur observers and map producers contributing to the historical overview in Fig. \ref{figS1}.  A more comprehensive historical discussion is given by \citet{17rogers}.}
 \centering
 \begin{tabular}{ccc}
 \hline
Date & Observer(s) & Map Producer \\
\hline
1995-Jul-4/5 & I. Miyazaki & H.J. Mettig \\
1996-Jul-19/22 & I. Miyazaki & H.J. Mettig \\
1997-Jul-11/14 & I. Miyazaki & H.J. Mettig \\
1998-Sep-24/26 & I. Miyazaki & H.J. Mettig \\
1999-Nov-10/11 & I. Miyazaki & H.J. Mettig \\
2002-Jan-18/21 & A. Cidadao, D. Parker & H.J. Mettig \\
2003-Jan-28/29 & D. Peach & D. Peach \\
2004-Mar-18/19 & D. Parker & D. Peach \\
2005-Apr-25/26 & D. Peach & D. Peach \\
2006-Apr-19/20 & D. Peach & D. Peach \\
2007-May-25/27 & D. Peach & D. Peach \\
2008-May-9/10 & P. Haese & M. Jacquesson \\
2009-Sep-5/6 & D. Peach & D. Peach \\
2010-Sep-12/13 & D. Peach & D. Peach \\
2011-Nov-6 & W. Jaeschke & W. Jaeschke \\
2012-Aug-20 & C. Go & M. Vedovato \\
2013-Mar-20 & D. Parker & M. Vedovato \\
2014-Feb-19 & M. Kardasis & M. Kardasis \\
2015-Feb-7& K. Quin & K. Quin \\
 \hline
 \end{tabular}
 \end{table}

\acknowledgments
All data can be obtained from the primary author (LNF, email: leigh.fletcher@leicester.ac.uk) and the relevant observatory archives.  We are extremely grateful to all those participating in the Earth-based support campaign for the Juno mission.  Fletcher was supported by a Royal Society Research Fellowship at the University of Leicester.  The UK authors acknowledge the support of the Science and Technology Facilities Council (STFC).  A portion of this work was performed by Orton at the Jet Propulsion Laboratory, California Institute of Technology, under a contract with NASA.  This research used the ALICE High Performance Computing Facility at the University of Leicester.  

This investigation was partially based on thermal-infrared observations acquired at (i) the ESO Very Large Telescope Paranal UT3/Melipal Observatory (program IDs 096.C-0091 and 097.C-0222); (ii) Subaru Telescope and obtained from the SMOKA database, which is operated by the Astronomy Data Center, National Astronomical Observatory of Japan (program ID S16B-025); and (iii) NASA's Infrared Telescope Facility, which is operated by the University of Hawaii under contract NNH14CK55B with the National Aeronautics and Space Administration (program IDs 2015B018 and 2016A027).  Raw data from each of these programs will be available via the observatory archives, reduced data are available from the primary author.  Hubble Space Telescope WFC3 maps of Jupiter were acquired as part of the OPAL program (Credit:  A. Simon, M. Wong, G. Orton), and available via the Mikulski Archive for Space Telescopes (https://archive.stsci.edu/prepds/opal/).  IRTF SPeX observations were acquired by G. Orton and T. Momary and are available via http://junoirtf.space.swri.edu. We are extremely grateful to M. Vedovato for providing cylindrical reprojections of visible-light observations by C. Go and H. Einaga (9-11 June 2016) in Fig. 1, and to J. Tollefson for providing the zonal wind profiles from Hubble OPAL data. We wish to recognize and acknowledge the significant cultural role and reverence that the summit of Mauna Kea has always had within the indigenous Hawaiian community.  We are most fortunate to have the opportunity to conduct observations from this mountain.

\bibliography{references}




\listofchanges

\end{document}